\documentclass[12pt,a4paper]{article}
\pdfoutput=1
\usepackage[utf8]{inputenc}
\usepackage[T1]{fontenc}
\usepackage{lmodern}
\usepackage[british]{babel}
\usepackage{amsmath,slashed}
\usepackage{amssymb}
\usepackage{amsfonts}
\usepackage{cite}
\usepackage{color}
\usepackage{float}
\usepackage{stmaryrd}
\usepackage{url}
\usepackage[colorlinks=true, linkcolor=blue, urlcolor=blue, citecolor=blue, anchorcolor=blue]{hyperref}
\usepackage{graphicx,psfrag}
\usepackage{placeins,bbm}
\usepackage{subfigure}
\usepackage{todonotes}
\usepackage[margin=3cm]{geometry}
\usepackage{authblk}

\newcommand{\Dtilde}{\ensuremath{\widetilde D} }
\newcommand{\Ddag}{\ensuremath{D^{\dag}} }
\newcommand{\cD}{\ensuremath{\mathcal D} }
\newcommand{\cDbar}{\ensuremath{\overline{\mathcal D}} }

\newcommand{\cN}{\ensuremath{\mathcal N} }
\newcommand{\cO}{\ensuremath{\mathcal O} }
\newcommand{\cP}{\ensuremath{\mathcal P} }
\newcommand{\cQ}{\ensuremath{\mathcal Q} }
\newcommand{\cU}{\ensuremath{\mathcal U} }
\newcommand{\cUbar}{\ensuremath{\overline{\mathcal U}} }
\newcommand{\Ibb}{\ensuremath{\mathbb I} }
\newcommand{\Pbb}{\ensuremath{\mathbb P} }
\newcommand{\Xbb}{\ensuremath{\mathbb X} }
\newcommand{\be}{\ensuremath{\beta} }
\newcommand{\ga}{\ensuremath{\gamma} }
\newcommand{\gaEff}{\ensuremath{\ga_{\text{eff}}} }
\newcommand{\de}{\ensuremath{\delta} }
\newcommand{\vareps}{\ensuremath{\varepsilon} }
\newcommand{\La}{\ensuremath{\Lambda} }
\newcommand{\la}{\ensuremath{\lambda} }
\newcommand{\lalat}{\ensuremath{\la_{\text{lat}}} }
\newcommand{\latilde}{\ensuremath{\widetilde{\la}} }
\newcommand{\lamin}{\ensuremath{\latilde_{\text{min}}^2} }
\newcommand{\lamax}{\ensuremath{\latilde_{\text{max}}^2} }
\newcommand{\muhat}{\ensuremath{\widehat\mu} }
\newcommand{\Om}{\ensuremath{\Omega} }
\newcommand{\om}{\ensuremath{\omega} }
\newcommand{\chitilde}{\ensuremath{\widetilde \chi} }
\newcommand{\rhohat}{\ensuremath{\widehat \rho} }
\newcommand{\Ombar}{\ensuremath{\overline \Om} }
\newcommand{\chidof}{\ensuremath{\chi^2 / \text{d.o.f.}} }
\newcommand{\SO}[1]{\ensuremath{\text{SO(}#1\text{)}} }
\newcommand{\SU}[1]{\ensuremath{\text{SU(}#1\text{)}} }
\newcommand{\U}[1]{\ensuremath{\text{U(}#1\text{)}} }
\newcommand{\Uone}{\ensuremath{\text{U(1)}} }
\newcommand{\X}{\ensuremath{\!\cdot\!} }
\newcommand{\lra}{\ensuremath{\longrightarrow} }
\DeclareMathOperator{\Tr}{Tr}
\newcommand{\vev}[1]{\ensuremath{\left\langle #1 \right\rangle} }
\renewcommand{\eqref}[1]{Eq.~(\ref{#1})}
\newcommand{\fig}[1]{Figure~\ref{#1}}
\newcommand{\tab}[1]{Table~\ref{#1}}
\newcommand{\refcite}[1]{Ref.~\cite{#1}}
\newcommand{\secref}[1]{Section~\ref{#1}}
\title{Eigenvalue spectrum and scaling dimension \\ of lattice $\cN = 4$ supersymmetric Yang--Mills}
\author[1]{Georg Bergner\thanks{georg.bergner@uni-jena.de}}
\author[2]{David Schaich\thanks{david.schaich@liverpool.ac.uk}}
\affil[1]{University of Jena, Institute for Theoretical Physics, Max-Wien-Platz 1, D-07743 Jena, Germany}
\affil[2]{Department of Mathematical Sciences, University of Liverpool, Liverpool L69 7ZL, United Kingdom}
\date{} % Suppress compilation date in frontmatter

\begin{document}
\maketitle

\begin{abstract}
  We investigate the lattice regularization of $\cN = 4$ supersymmetric Yang--Mills theory, by stochastically computing the eigenvalue mode number of the fermion operator.
  This provides important insight into the non-perturbative renormalization group flow of the lattice theory, through the definition of a scale-dependent effective mass anomalous dimension.
  While this anomalous dimension is expected to vanish in the conformal continuum theory, the finite lattice volume and lattice spacing generically lead to non-zero values, which we use to study the approach to the continuum limit.
  Our numerical results, comparing multiple lattice volumes, 't~Hooft couplings, and numbers of colors, confirm convergence towards the expected continuum result, while quantifying the increasing significance of lattice artifacts at larger couplings.
\end{abstract}

\section{\label{sec:intro}Introduction}
Four-dimensional maximally supersymmetric Yang--Mills theory ($\cN = 4$ SYM) is widely studied in theoretical physics.
Its many symmetries---including conformal symmetry and an SU(4) R-symmetry in addition to $Q = 16$ supersymmetries---make it arguably one of the simplest non-trivial quantum field theories in four dimensions, especially in the large-$N_c$ planar limit of its SU($N_c$) gauge group~\cite{ArkaniHamed:2008gz}.
This simplicity enabled its role as the conformal field theory of the original AdS/CFT holographic duality~\cite{Maldacena:1997re}, provided early insight into S-duality~\cite{Osborn:1979tq}, and continues to inform modern analyses of scattering amplitudes~\cite{Elvang:2015rqa}.

At the same time, the non-triviality of $\cN = 4$ SYM makes it important to explore the lattice regularization of the theory.
In addition to providing in principle a non-perturbative definition of $\cN = 4$ SYM, lattice field theory is also a way to numerically predict its behavior from first principles, even at strong coupling and away from the planar limit.
A prominent target for such predictions is the spectrum of conformal scaling dimensions, which depend on the 't~Hooft coupling $\la = N g_{\text{YM}}^2$.

Although lattice field theory has been very successfully used to analyze non-supersymmetric vector-like gauge theories such as quantum chromodynamics (QCD), it has proven more challenging to apply this approach to supersymmetric systems.
In large part this is because supersymmetry is explicitly broken by the lattice discretization of space-time.
See Refs.~\cite{Catterall:2009it, Bergner:2016sbv, Schaich:2018mmv} for reviews of these difficulties and the significant progress that has been achieved to overcome them in recent years.

In particular, using ideas borrowed from topological field theory and orbifold constructions, a lattice formulation of $\cN = 4$ SYM has been developed which preserves a closed supersymmetry subalgebra at non-zero lattice spacing $a > 0$~\cite{Catterall:2009it, Bergner:2016sbv, Schaich:2018mmv}.
Although 15 of the 16 supersymmetries are still broken away from the $a \to 0$ continuum limit, the single preserved supersymmetry significantly simplifies the lattice theory.
These simplifications are sufficient to establish that at most a single marginal coupling may need to be tuned to correctly recover the full symmetries of $\cN = 4$ SYM in the continuum limit~\cite{Catterall:2013roa, Catterall:2014mha}.
In addition, the moduli space of the lattice theory matches that of continuum $\cN = 4$ SYM to all orders in perturbation theory, and the renormalization group (RG) \be function vanishes at one loop in lattice perturbation theory~\cite{Catterall:2011pd}.

Of course, numerical lattice field theory calculations require both a non-zero lattice spacing that corresponds to a ultraviolet (UV) cutoff scale $1 / a$, as well as a finite lattice volume $(L\!\cdot\!a)^4$ that introduces an effective infrared (IR) cutoff, explicitly breaking conformal scale invariance.
It is also necessary to softly break the single preserved supersymmetry in order to regulate flat directions and make the lattice path integral well defined~\cite{Schaich:2018mmv, Catterall:2015ira}.
These facts make it challenging to analyze the approach to the $a \to 0$ continuum limit in the strongly interacting regime where lattice perturbation theory is unreliable.\footnote{The vanishing \be function in the continuum limit makes this problem much more difficult than the case of lattice QCD, where asymptotic freedom can guarantee weak coupling at the UV scale of the lattice spacing, even when the lattice volume is large enough to access strong coupling at long distances.}
It is therefore essential to carry out detailed numerical studies of the non-perturbative RG properties of the lattice theory.

In this paper we present progress investigating RG properties of lattice $\cN = 4$ SYM.
Specifically, we compute the eigenvalue mode number of the fermion operator, and use this to estimate the `mass anomalous dimension' $\ga^*(\la)$ that would appear in the scaling dimension of the corresponding fermion bilinear.
In the continuum theory this anomalous dimension is expected to vanish, $\ga^* = 0$, for all values of the 't~Hooft coupling $\la$. % Elaborate on why?
Computing $\ga^*(\la)$ is hence a way to assess the effects of breaking supersymmetry and conformal symmetry in numerical lattice calculations, and verify that the properties of $\cN = 4$ SYM are correctly reproduced in the continuum limit.
\refcite{Weir:2013zua} presented a first preliminary investigation of this topic, and preliminary results from a similar project studying the anomalous dimension of the Konishi operator more recently appeared in \refcite{Schaich:2018mmv}.
While \refcite{Weir:2013zua} proceeded by numerically computing the fermion operator eigenvalues, that approach quickly becomes inefficient as the lattice volume increases.
Here we instead apply stochastic techniques to estimate the mode number.

After reviewing the basic features of lattice $\cN = 4$ SYM in the next section, we summarize these stochastic techniques in \secref{sec:mode}, also discussing how the resulting mode number provides information on the anomalous dimension.
In \secref{sec:free} we consider the free ($\la = 0$) lattice theory, both to check our methods and to explore discretization artifacts.
Our numerical results for both the low-lying eigenvalues and the stochastic mode number are presented in \secref{sec:results}, and in \secref{sec:gamma} we use these to estimate the anomalous dimension.
After looking more closely at the dependence of the results on the gauge group and lattice volume in \secref{sec:NcL}, we conclude in \secref{sec:conc} with some discussion of the next steps for lattice analyses of $\cN = 4$ SYM.

\section{\label{sec:action}Lattice formulation of twisted $\cN = 4$ SYM}
As mentioned above, the lattice formulation of $\cN = 4$ SYM that we use has its origins in both topologically twisted~\cite{Sugino:2003yb, Sugino:2004qd, Catterall:2004np} and orbifolded~\cite{Cohen:2003xe, Cohen:2003qw, Kaplan:2005ta, Unsal:2006qp} approaches, which ultimately produce equivalent constructions~\cite{Catterall:2007kn, Damgaard:2008pa}.
Here we will use the twisted language, which organizes the $Q = 16$ supercharges of the theory into integer-spin representations of a twisted rotation group $\SO{4}_{\text{tw}} \equiv \mbox{diag}\left[\SO{4}_{\textrm{euc}} \otimes \SO{4}_R\right]$, where $\SO{4}_{\textrm{euc}}$ is the Lorentz group Wick-rotated to euclidean space-time and $\SO{4}_R$ is a subgroup of the SU(4) R-symmetry.
It is then convenient to combine these representations into $1$-, $5$- and $10$-component sets $\cQ$, $\cQ_a$ and $\cQ_{ab} = -\cQ_{ba}$, respectively.
These transform under the $S_5$ point-group symmetry of the $A_4^*$ lattice we use to discretize space-time, which consists of five basis vectors symmetrically spanning four dimensions.
The twisted-scalar supersymmetry is nilpotent, preserving the subalgebra $\{\cQ, \cQ\} = 0$ even at non-zero lattice spacing where $\cQ_a$ and $\cQ_{ab}$ are broken.

The lattice action, just like the continuum theory, is now the sum of the following $\cQ$-exact and $\cQ$-closed terms~\cite{Catterall:2011pd, Catterall:2012yq, Catterall:2013roa, Catterall:2014vka, Catterall:2014mha, Schaich:2014pda, Catterall:2015ira}:
\begin{equation}
  \label{eq:action}
  \begin{split}
    S_{\text{exact}}  & = \frac{N}{4\lalat} \sum_n \Tr\left[\cQ \left(\chi_{ab}(n)\cD_a^{(+)}\cU_b(n) + \eta(n) \cDbar_a^{(-)}\cU_a(n) - \frac{1}{2}\eta(n) d(n) \right)\right] \\
    S_{\text{closed}} & = -\frac{N}{16\lalat} \sum_n \Tr\left[\vareps_{abcde}\ \chi_{de}(n + \muhat_a + \muhat_b + \muhat_c) \cDbar_c^{(-)} \chi_{ab}(n)\right],
  \end{split}
\end{equation}
where $n$ indexes the lattice sites and repeated indices are summed.
In this paper we will present results in terms of the input lattice 't~Hooft coupling $\lalat$, which differs slightly from the continuum $\la$~\cite{Catterall:2014vka, Schaich:2018mmv}. % Elaborate on why?
The fermion fields $\eta$, $\psi_a$ and $\chi_{ab} = -\chi_{ba}$ transform in the same way as the corresponding twisted supercharges, and are respectively associated with the lattice sites, links and oriented plaquettes.
The gauge and scalar fields are combined into the five-component complexified gauge links $\cU_a$ and $\cUbar_a$, which appear in the finite-difference operators $\cD_a^{(+)}$ and $\cDbar_a^{(-)}$~\cite{Catterall:2007kn, Damgaard:2008pa}.

These complexified gauge links imply $\U{N} = \SU{N}\times \Uone$ gauge invariance, with flat directions in both the SU($N$) and U(1) sectors that need to be regulated in numerical calculations, as mentioned in \secref{sec:intro}.
To achieve this, we work with the improved action introduced by \refcite{Catterall:2015ira},\footnote{There is ongoing exploration of alternative lattice actions that address this issue in different ways~\cite{Catterall:2020lsi}.} which adds two deformations to \eqref{eq:action}.
The first is a simple scalar potential with tunable parameter $\mu$,
\begin{equation}
  \label{eq:pot}
  S_{\text{scalar}} = \frac{N}{4\lalat} \mu^2 \sum_n \sum_a \left(\frac{1}{N} \Tr\left[\cU_a(n) \cUbar_a(n)\right] - 1\right)^2,
\end{equation}
which regulates the SU($N$) flat directions while softly breaking the \cQ supersymmetry.
The second deformation is $\cQ$-exact, and replaces the term
\begin{equation}
  \label{eq:det}
  \cQ \left(\eta(n) \cDbar_a^{(-)}\cU_a(n)\right) \lra \cQ \left(\eta(n) \left[\cDbar_a^{(-)}\cU_a(n) + G\sum_{a \neq b} \left(\det\cP_{ab}(n) - 1\right) \Ibb_{N_c}\right]\right)
\end{equation}
in $S_{\text{exact}}$, with tunable parameter $G$.
This deformation picks out the U(1) sector through the determinant of the plaquette oriented in the $a$--$b$ plane, $\cP_{ab}(n)$, which is an $N_c\times N_c$ matrix at each lattice site $n$.
Expanding the \cQ transformation produces terms that modify both the fermion operator and the bosonic action~\cite{Catterall:2015ira}, as expected for a supersymmetric deformation.

Using this improved action, we have generated many ensembles of field configurations using the rational hybrid Monte Carlo (RHMC) algorithm~\cite{Clark:2006fx} implemented in parallel software that we make publicly available~\cite{Schaich:2014pda}.\footnote{{\tt\href{https://github.com/daschaich/susy}{github.com/daschaich/susy}}}
In addition, we have modified this software to implement the stochastic estimation of the mode number discussed below.
For this stochastic computation, it is convenient to rescale some of the fermion field components, which is irrelevant for the path integral since it introduces only a constant prefactor.
This rescaling is done only for the measurement of the mode number, not yet in RHMC configuration generation.

The particular rescaling we perform is chosen to put the fermion operator into its most symmetric form, which simplifies analytic considerations and changes the degeneracies of eigenvalues.
Reference~\cite{Catterall:2011pd} previously reported on the analytic structure of the lattice theory.
In its conventions, which differ slightly from \eqref{eq:action}, the fermion operator \Dtilde has the form
\begin{equation} % Equation 6.18 of Catterall:2011pd
  \Psi^T \Dtilde \Psi = \chitilde_{ab} \cD^{(+)}_{[a} \psi_{b]} + \eta \cD^{\dag (-)}_a \psi_a + \frac{1}{2}\vareps_{abcde} \chitilde_{ab} \cD^{\dag (-)}_c \chitilde_{de},
\end{equation}
where $\Psi = (\eta, \psi_a, \chi_{ab})$ collects the $16$ fermion fields into a vector.
We adjust this operator by reducing summations for $\chitilde_{ab}$ to the relevant part over $a < b$, compensating a factor of $2$ by rescaling $\chi_{ab} = 2 \chitilde_{ab}$.
The same result up to an overall factor of two is obtained from \eqref{eq:action} by rescaling $\eta \to \frac{\eta}{2}$.

The more symmetric fermion operator $D$ defined in this way is equivalent to the original operator in \refcite{Kaplan:2005ta}, and a rescaling was also done to discuss symmetries in \refcite{Catterall:2013roa}.
In the free theory, the squared operator $\Ddag D$ is now block diagonal in momentum space, $\Ddag D \sim f(p) \Ibb_{16N_c^2}$, implying a $16N_c^2$-fold degeneracy of the eigenvalues.
The function $f(p)$ on the $A_4^*$ lattice is
\begin{equation}
  \label{eq:fofp}
  f(p) = 4\sum_{\mu = 1}^4 \sin^2\left(p_{\mu} / 2\right) + 4\sin^2\left(-\sum_{\mu = 1}^4 p_{\mu} / 2\right),
\end{equation}
where $p_{\mu}$ are any four of the five linearly dependent lattice momenta~\cite{Catterall:2011pd}.
This $16N_c^2$-fold degeneracy is lifted in the interacting theory, but for any 't~Hooft coupling $\lalat \geq 0$ the eigenvalues of the lattice fermion operator (with or without rescaling) occur in $+/-$ pairs, so that the non-negative eigenvalues of the squared operator are always $2$-fold degenerate.
% Not bothering to mention that these $+/-$ eigenvalues are purely imaginary due to the operator being skew-symmetric...

\section{\label{sec:mode}The eigenvalue spectrum and stochastic estimation of the mode number}
The mode number, which is the integrated eigenvalue density of the fermion operator, allows for a precise estimate of the mass anomalous dimension~\cite{Patella:2012da, Cheng:2013bca, Fodor:2014zca}.
On the lattice the most practical definition is obtained from the spectral density of the massless fermion operator $D$,
\begin{equation}
  \rho(\om) = \frac{1}{V} \sum_k \vev{\de(\om - \la_k)}.
\end{equation}
Here the eigenvalues $\la_k$ should not be confused with the 't~Hooft coupling $\la$.
The mode number $\nu(\Om)$ is defined to be the number of eigenvalues $\la_k^2$ of the non-negative operator $\Ddag D$ that are smaller than $\Om^2$:
\begin{equation}
  \label{eq:mode}
  \nu(\Om) = \int_0^{\Om^2} \rhohat(\om) d\om = 2\int_0^{\Om} \rho(\om) d\om \, ,
\end{equation}
where \rhohat is the spectral density of $\Ddag D$ and the second equality follows from the eigenvalue pairing mentioned above.\footnote{Lattice QCD experts may expect the upper limit of integration in \eqref{eq:mode} to involve $\La = \sqrt{\Om^2 - m_R^2}$, with $m_R$ a renormalised fermion mass~\cite{Patella:2012da}.  In this work $m_R = 0$ and $\La = \Om$.}
Throughout the paper quantities like the eigenvalues $\la_k$ and the scale \Om are provided in lattice units.

The anomalous dimension $\ga^*$ governs the dependence of the mode number on the scale $\Om^2$:
\begin{equation}
  \label{eq:modenumber}
  \nu(\Om) \propto (\Om^2)^{2 / (1 + \ga^*)} \; .
\end{equation}
Additional terms present in the Wilson-fermion case (see \refcite{Patella:2012da}) do not appear here.
This makes it possible to define a scale-dependent effective anomalous dimension from any two values of the mode number:
\begin{equation}
  \gaEff(\Ombar) = 2\frac{\log(\Om_1^2) - \log(\Om_2^2)}{\log(\nu(\Om_1)) - \log(\nu(\Om_2))} - 1 \; ,
  \label{eq:modneff}
\end{equation}
where $\Ombar^2 \equiv (\Om_1^2 + \Om_2^2) / 2$.
In addition to depending on the choice of scales $\Om_1$ and $\Om_2$, the determination of \gaEff on any ensemble of lattice field configurations will be affected by lattice artifacts.
As we discuss in \secref{sec:free}, even the free theory with $\lalat = 0$ only recovers the continuum $\ga^* = 0$ after extrapolation to the continuum limit.

If $\Om_1$ and $\Om_2$ are close to each other and the lattice is coarse, the results of this naive method are quite unstable and fluctuate significantly.
We will show below that fits provide more stable results.
An alternative approach to improve stability is to normalize the mode number with respect to the free-theory $\nu_{\text{free}}$, using some fixed reference scale $\Om_1$,
\begin{equation}
  \gaEff(\Om) = \frac{\log(\nu_{\text{free}}(\Om)) - \log(\nu_{\text{free}}(\Om_1))}{\log(\nu(\Om)) - \log(\nu(\Om_1))} - 1 \; .
  \label{eq:modneff2}
\end{equation}

Now that we have seen how an effective anomalous dimension can be extracted from the mode number, we review our stochastic estimation of $\nu(\Om)$ using the well-established projection method proposed in \refcite{Giusti:2002sm}.
This method is based on a rational approximation of the projection operator \Pbb for eigenvalues in the region below a given threshold.
The mode number is then
\begin{equation}
  \nu(\Om) = \vev{\Tr \Pbb(\Om)} \, ,
  \label{eq:projectionmethod}
\end{equation}
where the trace is obtained by stochastic estimation.
The projection operator is approximated in terms of the step function $h(x)$ using
\begin{align}
  \Pbb(\Om) & \approx h(\Xbb)^4 &
  \Xbb & = 1 - \frac{2\Om_*^2}{\Ddag D + \Om_*^2} \, .
\end{align}
The parameter $\Om_* \approx \Om$ is adjusted to minimize the error of the approximation---see \refcite{Giusti:2002sm} for further details.

More recently, a different method based on a Chebyshev expansion of the spectral density $\rho$ has been proposed~\cite{Fodor:2016hke, Bergner:2016hip}.
In this Chebyshev expansion method, the spectrum is rescaled to the interval $[-1,1]$ by defining
\begin{equation}
  M = \frac{2\Ddag D - \la_{\text{max}}^2 - \la_{\text{min}}^2}{\la_{\text{max}}^2 - \la_{\text{min}}^2} \; ,
\end{equation}
where $\la_{\text{max}}^2$ and $\la_{\text{min}}^2$ are the maximal and minimal eigenvalues of $\Ddag D$.
We consider the integral of the spectral density $\rho_M$ of the rescaled operator $M$ multiplied by the $n$th term $T_n$ of a Chebyshev polynomial of order $N_p$,
\begin{align}
  c_n & = \int_{-1}^1 \rho_M(x) T_n(x) dx &
  0 & \leq n \leq N_p,
\end{align}
which we estimate stochastically using $N_s$ random $Z_4$ noise vectors $v_l$: % Mention polynomial here matrix valued?
\begin{equation}
  c_n \approx \frac{1}{N_s} \sum_{l = 1}^{N_s} \vev{v_l | T_n(M) | v_l}.
\end{equation}
Based on the orthogonality relations for the $T_n$, the spectral density $\rho_M$ is now approximated by
\begin{equation}
  \rho_M(x) \approx \frac{1}{\pi \sqrt{1 - x^2}} \sum_{n = 0}^{N_p} (2 - \de_{n0}) c_n T_n(x).
\end{equation}
The spectral density of $\Ddag D$ is then obtained by mapping the interval $[-1, 1]$ back to the original eigenvalue region $[\la_{\text{min}}^2, \la_{\text{max}}^2]$, and can be integrated analytically to provide the mode number via \eqref{eq:mode}.

The work we present below will focus on measurements of the mode number using the Chebyshev expansion method.
We use polynomials of order $5{,}000 \leq N_p \leq 10{,}000$ depending on the spectral range $[\la_{\text{min}}^2, \la_{\text{max}}^2]$, which increases for stronger 't~Hooft couplings.
We cross-checked these results through the more computationally expensive projection method of \eqref{eq:projectionmethod}, as well as by directly computing the low-lying eigenvalues of $\Ddag D$ using a Davidson-type method provided by the PReconditioned Iterative Multi-Method Eigensolver (PRIMME) library~\cite{Stathopoulos:2010}.
In addition to checking the stochastic results for small $\Om$, these direct eigenvalue measurements can provide an alternative estimate of the effective anomalous dimension, from the volume-scaling relation~\cite{Weir:2013zua}
\begin{align}
  \vev{\la_k^2} \propto L^{-y_k} \; ,
  \label{eq:evscaling}
\end{align}
where $y_k = 2 / (1 + \gaEff)$ and $\vev{\la_k^2}$ is the average $k$th eigenvalue of $\Ddag D$.
We will see below that the mode number provides more precise results than the individual eigenvalues, as expected~\cite{Cheng:2013bca}.

\section{\label{sec:free}Discretization effects for the free lattice theory}
Before presenting numerical results obtained from analyzing the available ensembles of lattice $\cN = 4$ SYM field configurations, we consider the $\lalat = 0$ free theory to test both our methods of stochastically estimating the mode number, as well as our extraction of $\gaEff$.
By definition, all anomalous dimensions vanish for the free theory, meaning that any non-zero results we obtain will provide information about the lattice artifacts we want to explore.
The free theory is simple enough that we are able to analytically compute the mode number on the $A_4^*$ space-time lattice.

\begin{figure}
  \includegraphics[width=0.5\textwidth]{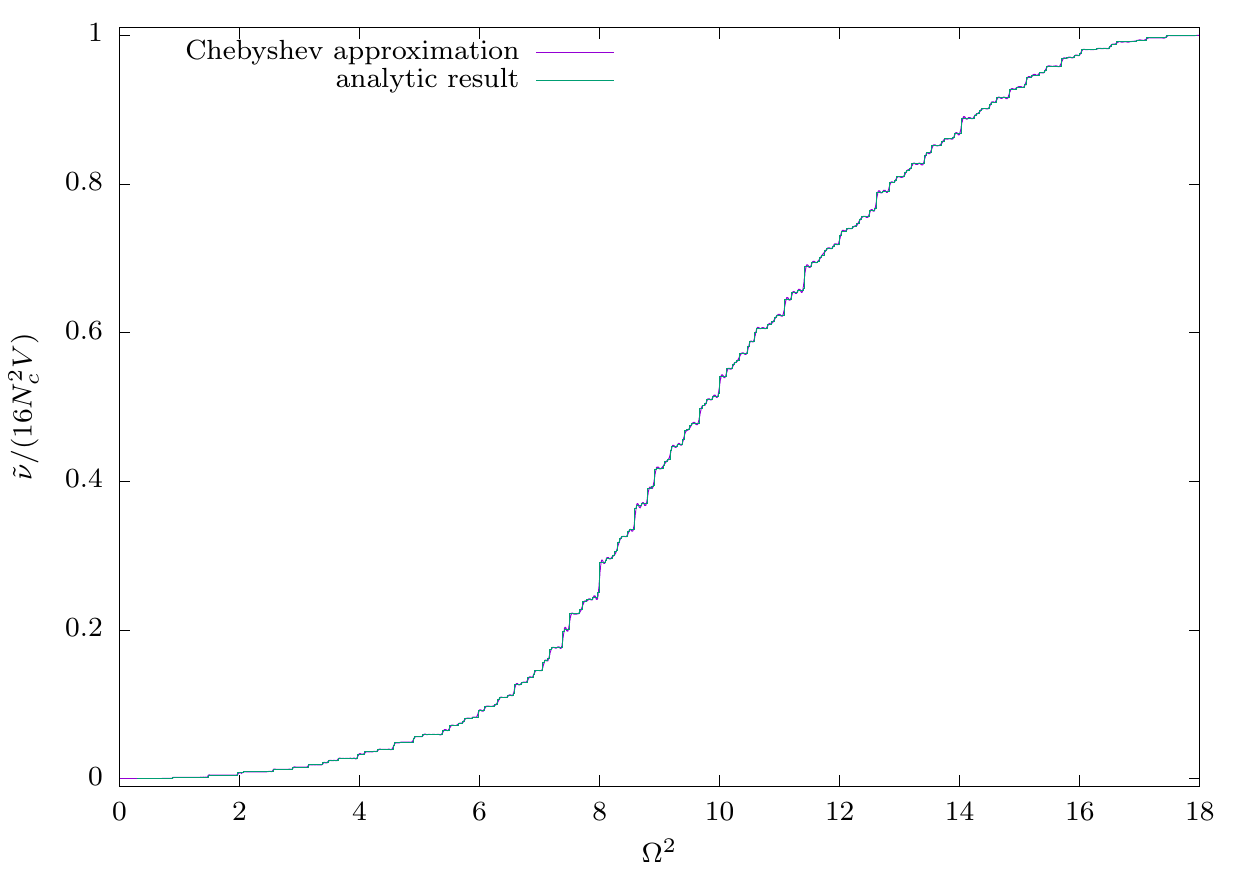}\hfill
  \includegraphics[width=0.5\textwidth]{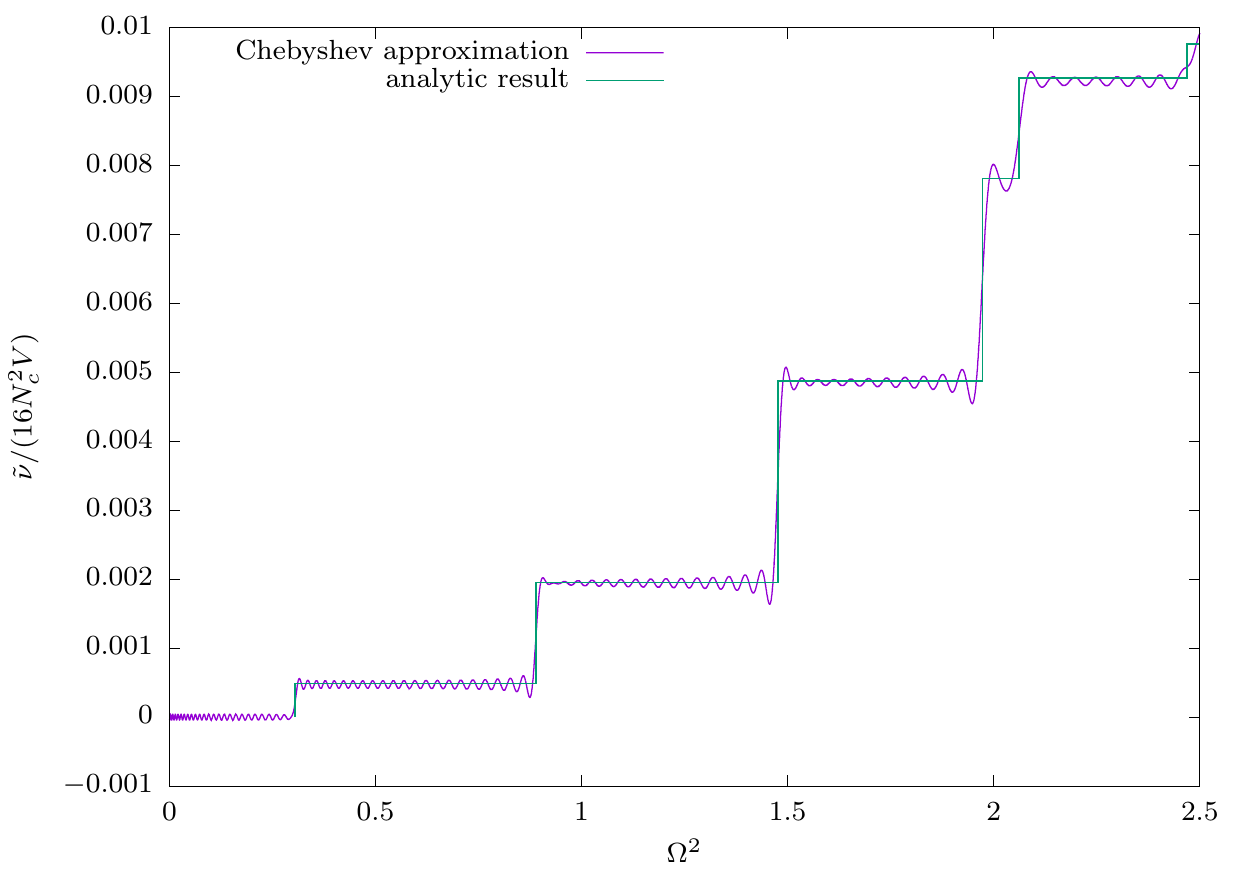}
  \caption{Free-theory mode number analytic result and Chebyshev approximation for lattice volume $V = 8^4$ with $N_c = 2$ and antiperiodic BCs, the latter using $N_p = 1000$ and $N_s = 10$ estimators.  The right plot zooms in on $\Om^2 \leq 2.5$ to show the stepwise behavior more clearly.  Both plots normalize the mode number by the total number of eigenvalues, $16N_c^2 V$.}
  \label{fig:chebfree}
\end{figure}

The free $\Ddag D$ operator has an enhanced symmetry, resulting in larger degeneracies of the eigenvalues, especially when all fields are subject to periodic boundary conditions (BCs) in all four dimensions.
In RHMC configuration generation, and in \fig{fig:chebfree}, antiperiodic BCs are applied to the fermion fields in the time direction, to lift a fermion zero mode.
Even with those antiperiodic BCs, many free-theory eigenvalues are degenerate, leading the mode number to exhibit distinct steps at certain values of $\Om$.
This stepwise behavior is typically quite difficult to capture with a polynomial approximation, but \fig{fig:chebfree} shows that the Chebyshev approach is able to provide reasonable precision.

\begin{figure}
  \centering
  \includegraphics[width=0.7\textwidth]{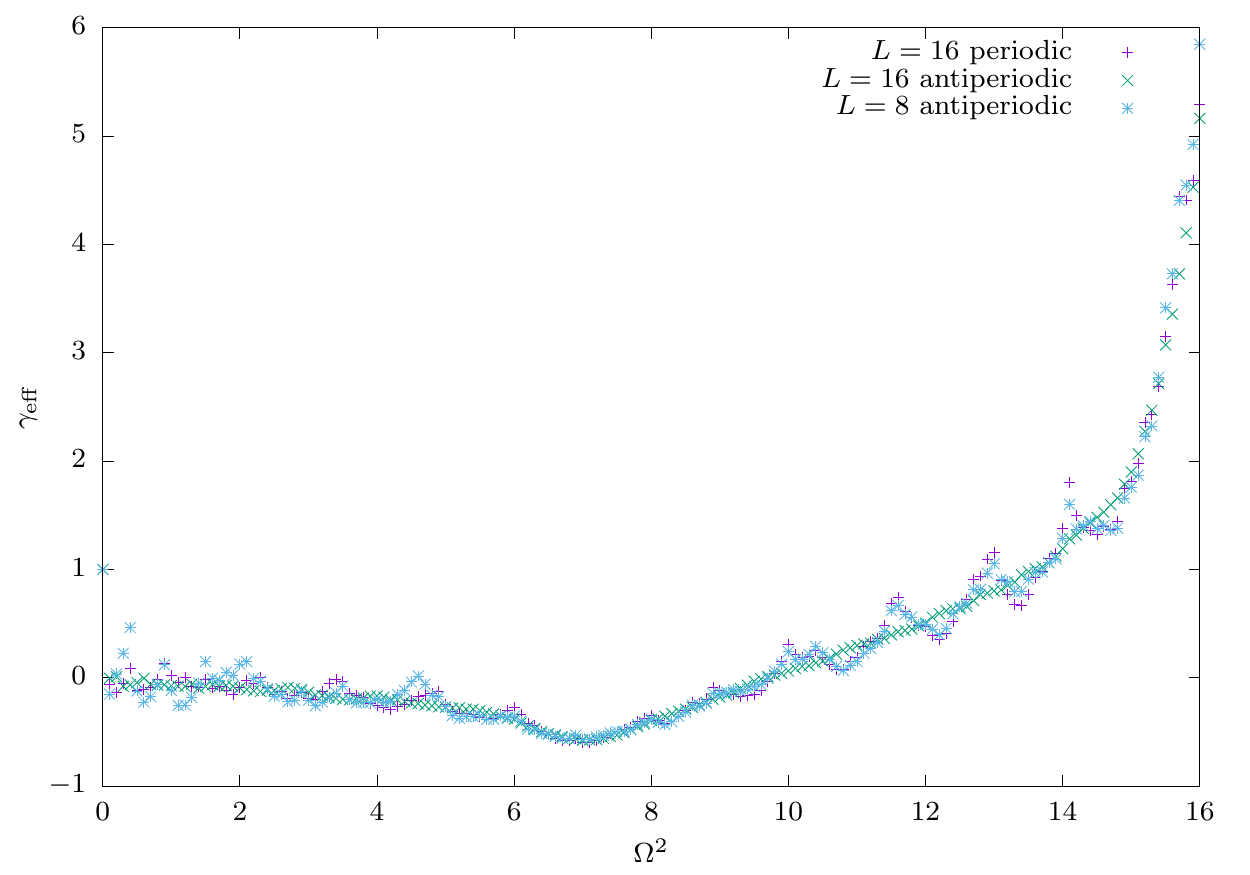}
  \caption{Effective anomalous dimension \gaEff obtained from using \eqref{eq:modenumber} to fit the analytic mode number of the free lattice $\Ddag D$ operator across the window $[\Om^2, \Om^2 + 1]$.  Two different $A_4^*$ lattice volumes $8^4$ and $16^4$ are compared, the latter considering both periodic and antiperiodic fermion BCs in the time direction.}
  \label{fig:fitgamma}
\end{figure}

While the effective anomalous dimension \gaEff can be estimated from any two values of the mode number using \eqref{eq:modneff}, this naive approach produces large fluctuations, especially for small \Om where the stepwise behavior of the mode number is most prominent.
More stable results are obtained by fitting the mode number according to \eqref{eq:modenumber} over a window $[\Om^2, \Om^2 + \ell]$ of length $\ell$, which we show in \fig{fig:fitgamma}.
This figure compares $\ell = 1$ fit results for the free-theory \gaEff from two different lattice volumes $8^4$ and $16^4$, considering both periodic and antiperiodic BCs for the latter.
As expected, we obtain more stable results as we approach the continuum limit by increasing the lattice volume.
We also see larger oscillations for the case of periodic BCs, which is due to the larger degeneracies of the eigenvalues in this case.

\begin{figure}
  \subfigure[$f(p)$]{\includegraphics[width=0.325\textwidth]{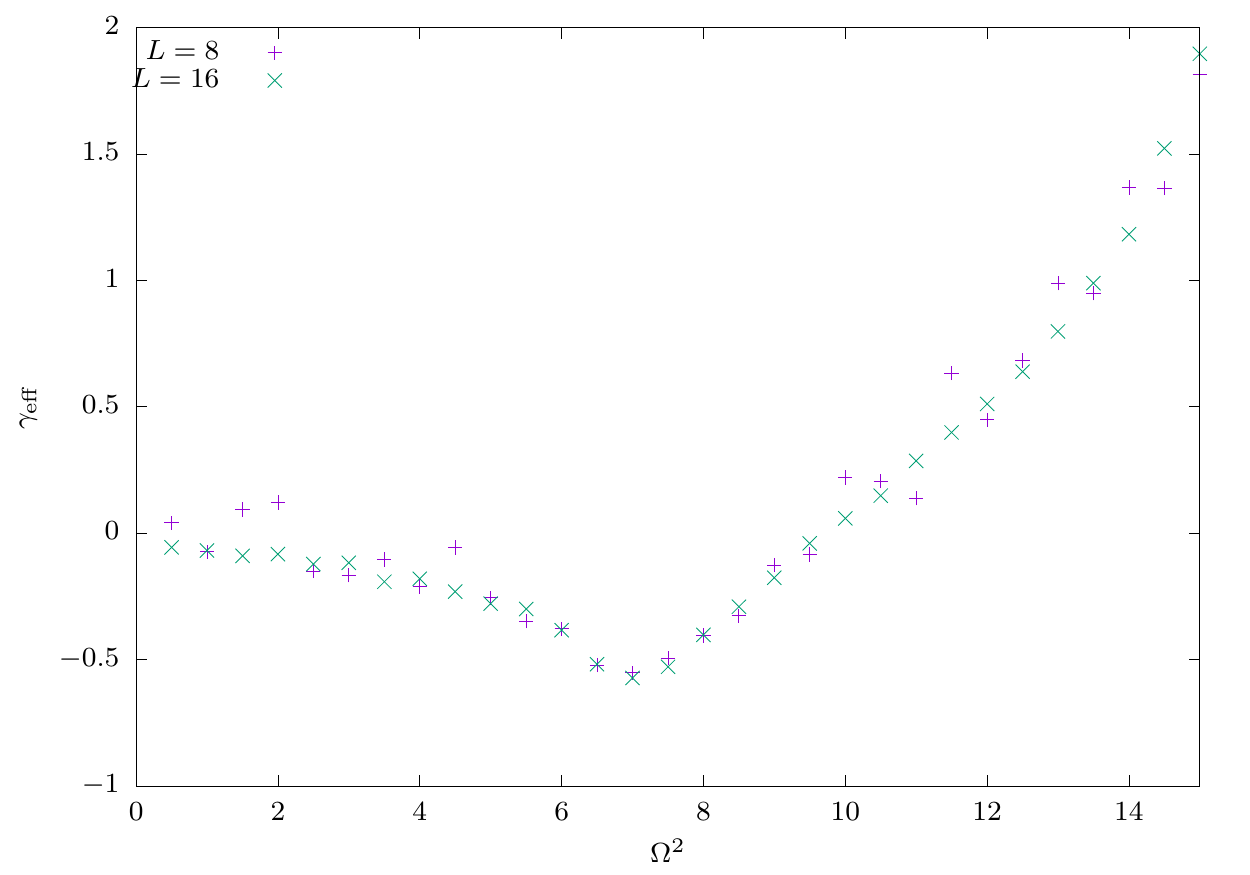}}\hfill
  \subfigure[$p^2$]{\includegraphics[width=0.325\textwidth]{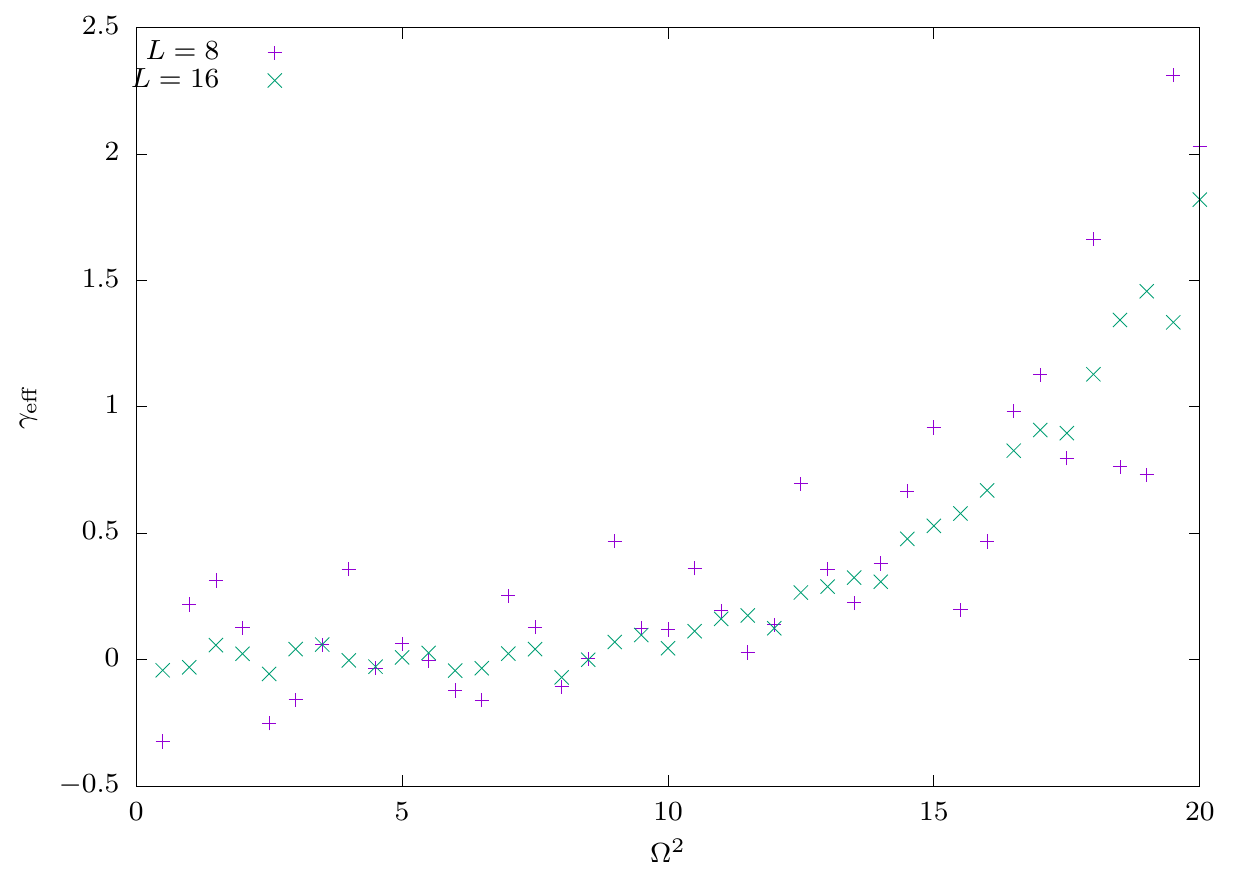}}\hfill
  \subfigure[$4\sin^2(p / 2)$]{\includegraphics[width=0.325\textwidth]{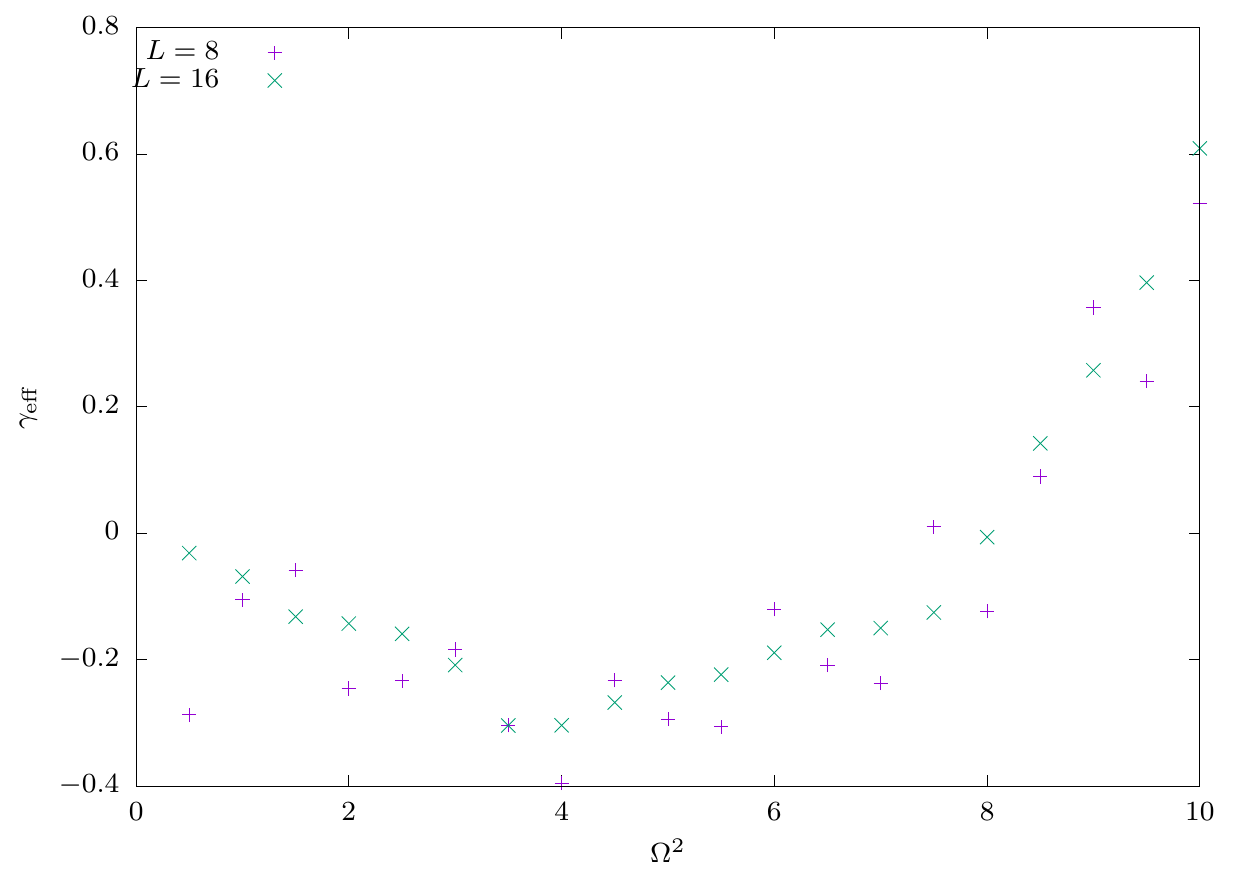}}
  \caption{Free-theory results for the effective anomalous dimension $\gaEff$ obtained from using \eqref{eq:modenumber} to fit the analytic mode number over the window $[\Om^2, \Om^2 + 1]$.  Each plot compares $L^4$ lattice volumes with $L = 8$ and $16$, using periodic BCs.  The left plot shows the $A_4^*$ lattice result using $f(p)$ from \eqref{eq:fofp}.  The center plot considers instead a naive continuum-like discretization of the free operator in momentum space, while the right plot corresponds to the hypercubic-lattice dispersion relation $4\sin^2(p / 2)$.}
  \label{fig:freedisc}
\end{figure}

While the IR limit $\lim_{\Om \to 0} \gaEff \approx 0$ for the free-theory results shown in \fig{fig:fitgamma}, as the scale increases up to $\Om \lesssim 7$, the effective anomalous dimension tends towards negative values.
In \fig{fig:freedisc} we confirm that this trend is due to lattice artifacts, by comparing three dispersion relations for the free operator in momentum space.
For the $A_4^*$ lattice with periodic BCs, this is based on the function $f(p)$ in \eqref{eq:fofp}.
Considering instead a continuum-like dispersion relation $p^2$ produces effective anomalous dimension results more consistent with the true $\ga^* = 0$, while the hypercubic-lattice dispersion relation $4\sin^2(p / 2)$ leads to the same qualitative trends in \gaEff as the $A_4^*$ lattice used for $\cN = 4$ SYM.

From these investigations of the free theory, we can conclude that the Chebyshev method provides a reasonable approximation of the mode number.
Even in this free case where the eigenvalue degeneracies are largest, the stepwise behavior of the mode number is accurately resolved.
The resulting oscillations in results for the effective anomalous dimension \gaEff are manageably small when we fit the mode number over a window of adequate length $\ell$, and are expected to decrease for the $\lalat > 0$ to which we now turn.

In addition to checking our methods, we can also use this consideration of the free theory to improve our main numerical analyses.
The ensembles of lattice $\cN = 4$ SYM field configurations we analyze span a range of 't~Hooft couplings $0.25 \leq \lalat \leq 2.5$ within which some similarities to the free theory persist.
Therefore we can employ the method of \eqref{eq:modneff2} as an alternative approach to improve the stability of the results by using the free theory as a reference for the scaling of the mode number.

\section{\label{sec:results}Results for eigenvalues and mode number}
\begin{table}
  \centering
  \renewcommand\arraystretch{1.2} % Increase the height of each row
  \addtolength{\tabcolsep}{2.5 pt}  % Increase separation between columns
  \begin{tabular}{c|c|l|l||c|c|l|c|r}
    $N_c$ & $L$ & \lalat & $\mu$ & \lamin       & \lamax & Spect.~range         & Meas. & $\tau(\lamin)$ \\
    \hline
    2     & 10  & 0.5    & 0.16  & $9\X10^{-3}$ & 25     & $[1\X10^{-4}, 45]$   & 700   &  4.1            \\
    \hline
    2     & 12  & 0.25   & 0.095 & $6\X10^{-3}$ & 23     & $[5\X10^{-5}, 2500]$ & 200   &  5.8            \\
    &     & 0.5    & 0.13  & $5\X10^{-3}$ & 25     & $[1\X10^{-7}, 1000]$ & 780   &  5.7            \\
    &     & 1.0    & 0.19  & $3\X10^{-3}$ & 29     & $[1\X10^{-3}, 50]$   & 240   &  2.6            \\
    &     & 1.5    & 0.23  & $2\X10^{-3}$ & 33     & $[5\X10^{-5}, 2500]$ & 300   &  3.2            \\
    &     & 2.0    & 0.25  & $1\X10^{-3}$ & 36     & $[5\X10^{-5}, 2500]$ & 200   &  2.9            \\
    &     & 2.5    & 0.3   & $7\X10^{-5}$ & 43     & $[1\X10^{-6}, 1900]$ & 190   & 10.1            \\
    \hline
    2     & 14  & 0.5    & 0.11  & $3\X10^{-3}$ & 25     & $[1\X10^{-4}, 45]$   & 700   & 10.3            \\
    \hline
    2     & 16  & 0.25   & 0.07  & $2\X10^{-3}$ & 22     & $[1\X10^{-5}, 50]$   & 180   & 11.3            \\
    &     & 0.5    & 0.1   & $2\X10^{-3}$ & 25     & $[1\X10^{-4}, 45]$   & 180   &  8.0            \\
    &     & 1.0    & 0.14  & $1\X10^{-3}$ & 29     & $[1\X10^{-5}, 50]$   & 410   &  4.2            \\
    &     & 1.5    & 0.17  & $6\X10^{-4}$ & 33     & $[1\X10^{-5}, 50]$   & 230   &  4.8            \\
    &     & 2.0    & 0.2   & $4\X10^{-4}$ & 38     & $[1\X10^{-5}, 50]$   & 250   &  7.5            \\
    &     & 2.5    & 0.22  & $2\X10^{-5}$ & 44     & $[1\X10^{-5}, 50]$   & 340   &  6.1            \\
    \hline
    3     & 12  & 0.5    & 0.15  & $3\X10^{-3}$ & 24     & $[5\X10^{-5}, 2500]$ & 150   &  3.7            \\
    &     & 1.0    & 0.2   & $1\X10^{-3}$ & 31     & $[5\X10^{-5}, 2500]$ & 200   &  2.2            \\
    &     & 1.5    & 0.23  & $1\X10^{-3}$ & 45     & $[5\X10^{-5}, 2500]$ & 150   &  2.2            \\
    \hline
    4     & 12  & 0.5    & 0.15  & $2\X10^{-3}$ & 23     & $[5\X10^{-5}, 2500]$ & 140   &  3.3            \\
  \end{tabular}
  \caption{\label{tab:ensembles}Summary of the ensembles we consider, with gauge group U($N_c$) and lattice volume $L^4$.  For all ensembles $G = 0.05$ and the fermion fields are subject to antiperiodic BCs in the time direction.  We set $\mu \approx \sqrt{5\lalat} / L$ to remove the scalar potential in \eqref{eq:pot} in the $L \to \infty$ continuum limit with fixed lattice 't~Hooft coupling $\lalat$.  For each ensemble we report the extremal eigenvalues $\latilde^2$ of $\Dtilde^{\dag} \Dtilde$, ensuring that they remain within the spectral range where the corresponding RHMC rational approximation is accurate.  The listed number of mode number measurements are carried out on thermalized configurations separated by $10$ molecular dynamics time units.  From measurements of \lamin on those same configurations we also estimate autocorrelation times $\tau$ in units of measurements.}
\end{table}

Our main numerical analyses involve the $18$ ensembles of lattice $\cN = 4$ SYM field configurations listed in \tab{tab:ensembles}, which are a subset of a broader collection of ensembles being used to investigate other aspects of the theory~\cite{Schaich:2018mmv}.
These were generated using the RHMC algorithm and the improved lattice action described in \secref{sec:action}.
For gauge group U(2) we consider $L^4$ lattice volumes with $10 \leq L \leq 16$, and lattice 't~Hooft couplings $0.25 \leq \lalat \leq 2.5$.
In addition we also analyze more limited data sets for gauge groups U(3) and U(4), which involve significantly larger computational costs.

As part of the process of configuration generation, we use PRIMME~\cite{Stathopoulos:2010} to measure the extremal eigenvalues of $\Dtilde^{\dag} \Dtilde$ on every saved configuration, ensuring that the minimum and maximum across the entire ensemble remain within the spectral range where the RHMC rational approximation is accurate.
This information is included in \tab{tab:ensembles}, where we also report the autocorrelation time of the smallest $\Dtilde^{\dag} \Dtilde$ eigenvalue following thermalization/equilibration, estimated using the `autocorr' module in \texttt{emcee}~\cite{Foreman:2013mc}.
This autocorrelation time provides an indication of how many statistically independent samples are available for each ensemble.
Some additional information about these ensembles is collected in the Appendix.

Following configuration generation and analysis of thermalization, we additionally compute the extremal eigenvalues of the more symmetric rescaled operator $\Ddag D$ on all thermalized configurations, each separated by $10$ molecular dynamics time units.
We also use all of these thermalized configurations to stochastically estimate the $\Ddag D$ mode number through the Chebyshev expansion method, with less-extensive cross-checks using the projection method, as described in \secref{sec:mode}.
Table~\ref{tab:ensembles} reports the number of measured configurations for each ensemble.

\begin{figure}
  \centering
  \includegraphics[width=0.6\textwidth]{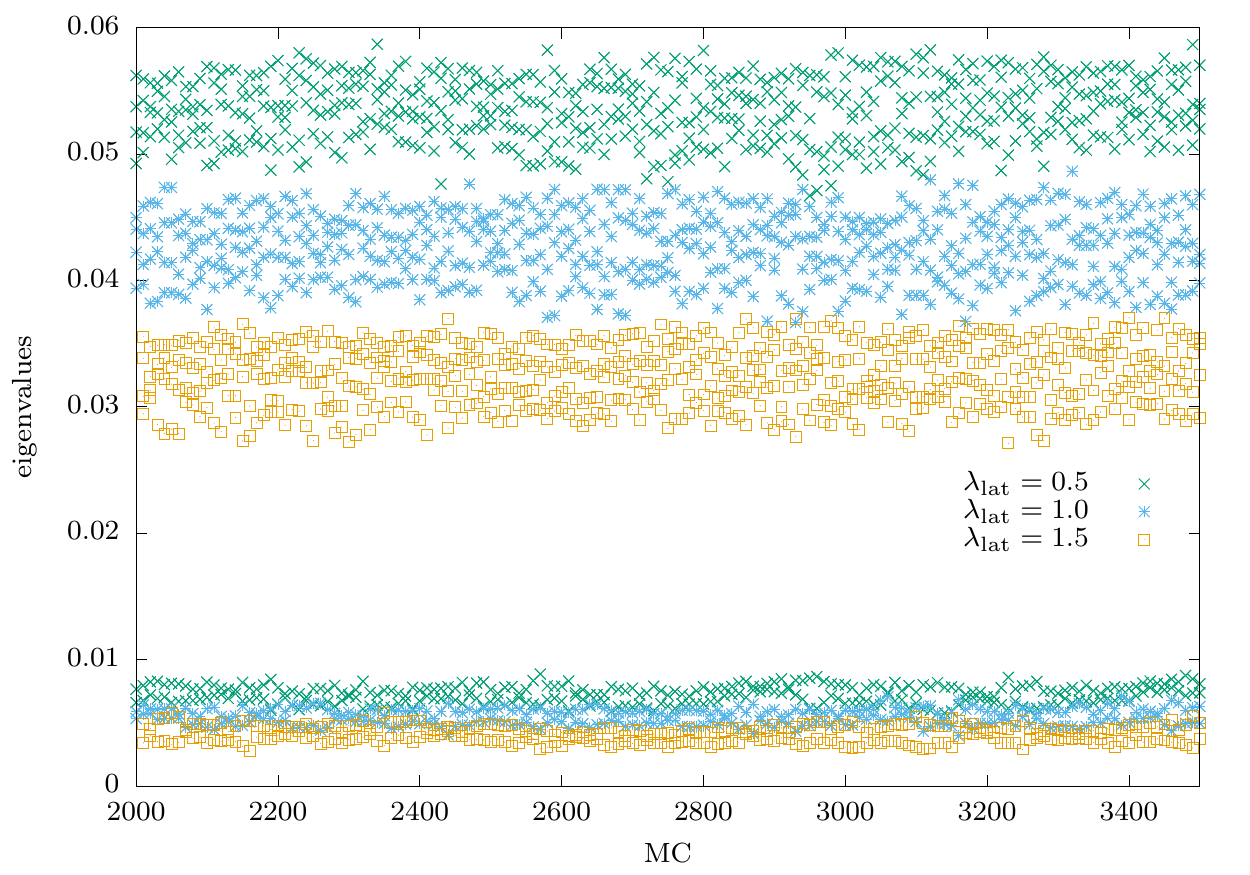}
  \caption{Monte Carlo history of the $2\times 6$ lowest $\Ddag D$ eigenvalue pairs for gauge group U(2) on $12^4$ lattices with different values of the 't~Hooft coupling $\lalat$.  The eigenvalues are measured on thermalized configurations separated by $10$ `MC' units on the horizontal axis.}
  \label{fig:eigenvals}
\end{figure}

The low-lying eigenvalues of $\Ddag D$ provide a first look at the mode number and its scaling.
Figure~\ref{fig:eigenvals} shows the general form of these low-lying eigenvalues, which features a pronounced gap between the lowest two pairs and the rest of the spectrum.
This gap is rather stable across each RHMC Markov chain, and its relative size decreases for smaller \lalat due to the lowest two pairs moving to larger values.
Similar observations were also made in \refcite{Weir:2013zua} for a different $\cN = 4$ SYM lattice action.
% Averages from the command below give 0.050/0.0067~7.5 for lalat=0.5, 0.039/0.0050~7.8 for lalat=1 and 0.029/0.0036~8.0 for lalat=1.5
% $ for i in l0.5_b0.13_G0.05 l1.0_b0.19_G0.05 l1.5_b0.23_G0.05 ; do grep "EIGENVALUE 4 " $i/Out/rescaled_eig.* | awk '{print $3}' > TEMP ; average TEMP ; grep "EIGENVALUE 0 " $i/Out/rescaled_eig.* | awk '{print $3}' > TEMP ; average TEMP ; rm TEMP ; done

\begin{figure}
  \centering
  \includegraphics[width=0.6\textwidth]{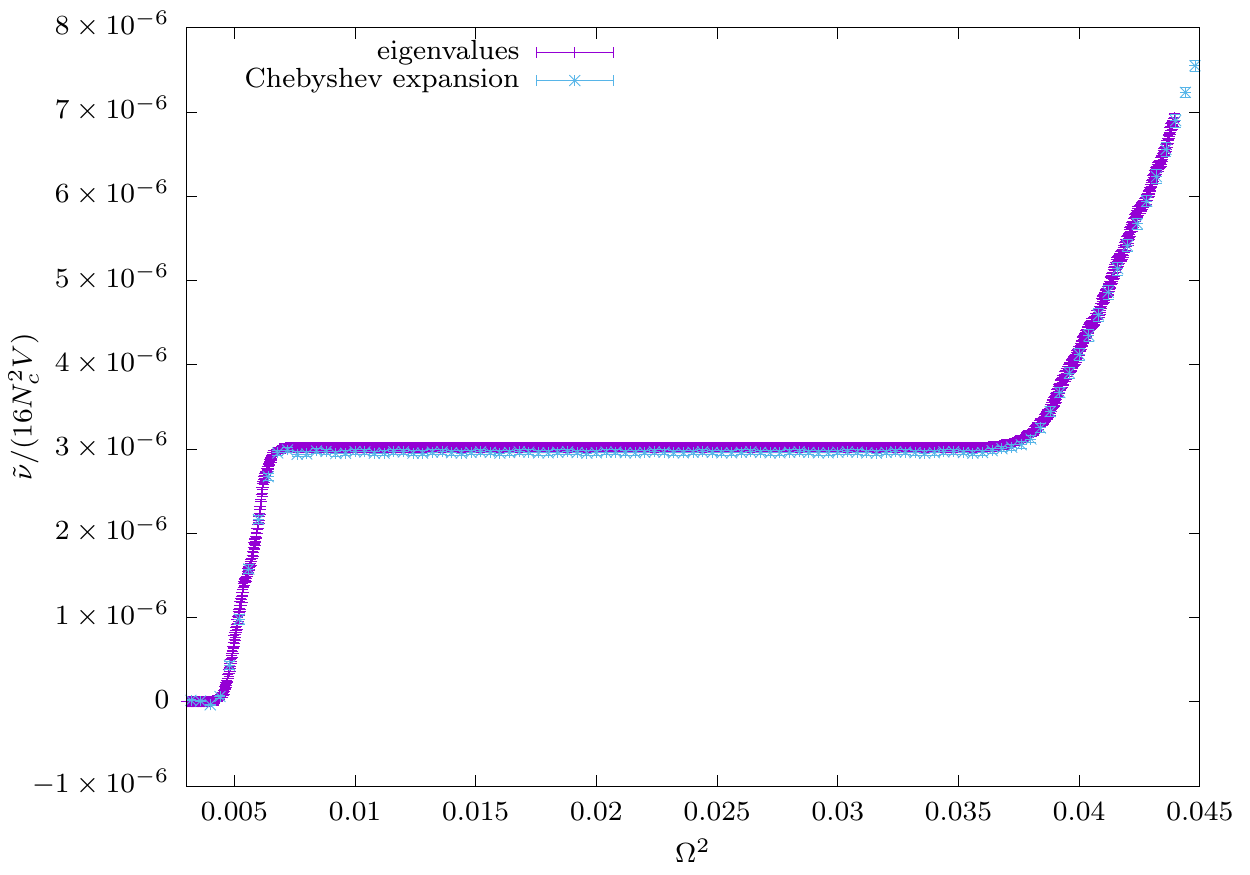}
  \caption{Normalized mode number obtained by two different methods on a $V = 12^4$ lattice for gauge group U(2) at $\lalat = 1$, averaging over all 240 measured configurations.  One method is the direct computation of the $2\times 6$ lowest $\Ddag D$ eigenvalue pairs shown in \fig{fig:eigenvals}.  The other method is a stochastic estimation of the Chebyshev expansion with $N_p = 5000$.}
  \label{fig:evcheb}
\end{figure}

From these eigenvalues we can directly compute the mode number for small $\Om$, allowing a first cross-check of the its stochastic estimation through the Chebyshev approach.
Figure~\ref{fig:evcheb} shows a representative example confirming that the Chebyshev method indeed reproduces all the features of the mode number obtained from the eigenvalue spectrum.
In particular, the large gap in the eigenvalue spectrum is clearly reflected by the plateau in the mode number.

\begin{figure}
  \centering
  \includegraphics[width=0.8\textwidth]{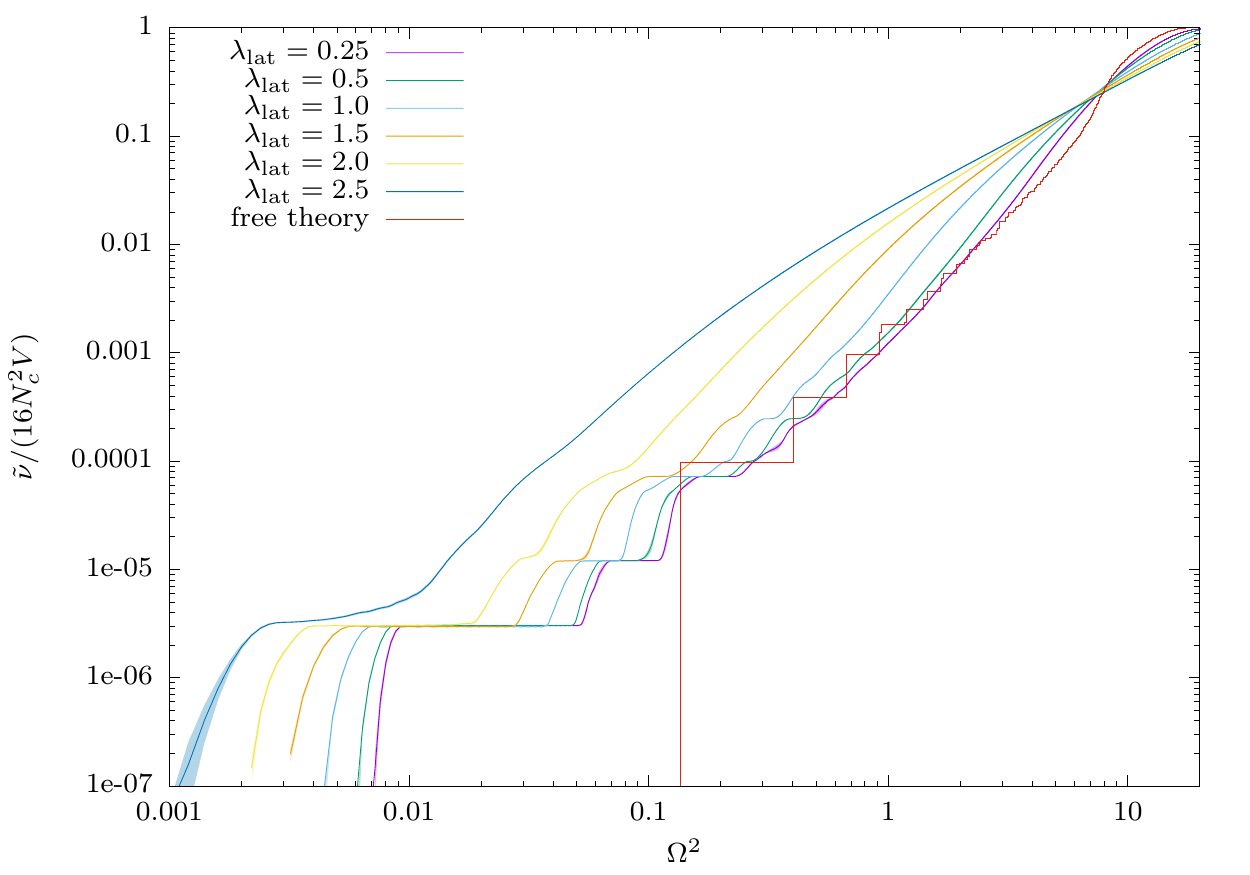}
  \caption{Normalized mode number obtained for gauge group U(2) on $12^4$ lattices for all six values of the coupling $\lalat$, on double-logarithmic axes.  We average over all the stochastic Chebyshev measurements specified in \tab{tab:ensembles} and include small errorbands showing the resulting statistical uncertainty.  The analytic result for the free theory is also shown.}
  \label{fig:all12}
\end{figure}

In \fig{fig:all12} we compare the mode number for all six available values of the lattice 't~Hooft coupling \lalat for $12^4$ lattices with gauge group U(2).
We include the analytic result for the free theory, and can see that the large degeneracies of the free-theory eigenvalues are broken even for the smallest $\lalat = 0.25$ we consider.
As the coupling gets stronger, the smallest eigenvalues move towards zero while larger fluctuations make the mode number a smoother function of $\Om$.
We now turn to the extraction of the effective anomalous dimension from these numerical results for the lattice $\cN = 4$ SYM mode number.

\section{\label{sec:gamma}Estimates for the anomalous dimension}
From \eqref{eq:modenumber} we see that the effective anomalous dimension appears in the slope $2 / (1 + \gaEff)$ of the mode number vs.\ $\Om^2$ on double-logarithmic axes, as shown in \fig{fig:all12}.
The general behavior shown in this figure therefore already reveals the main features of our results for \gaEff and its dependence on $\lalat$.
As we saw for the free theory in \secref{sec:free}, the stepwise behavior of the mode number at very small \Om obstructs precise extraction of $\gaEff$.
For larger \Om the slope decreases with increasing $\lalat$, which implies a larger $\gaEff$.
In addition, for stronger couplings this region of larger \gaEff extends towards smaller $\Om^2$, though the non-linearities in the results imply that \gaEff decreases for small \Om and may become consistent with the IR limit $\lim_{\Om \to 0} \gaEff \approx 0$ observed for the free theory and expected for the interacting continuum theory.

\begin{figure}
  \centering
  \includegraphics[width=0.8\textwidth]{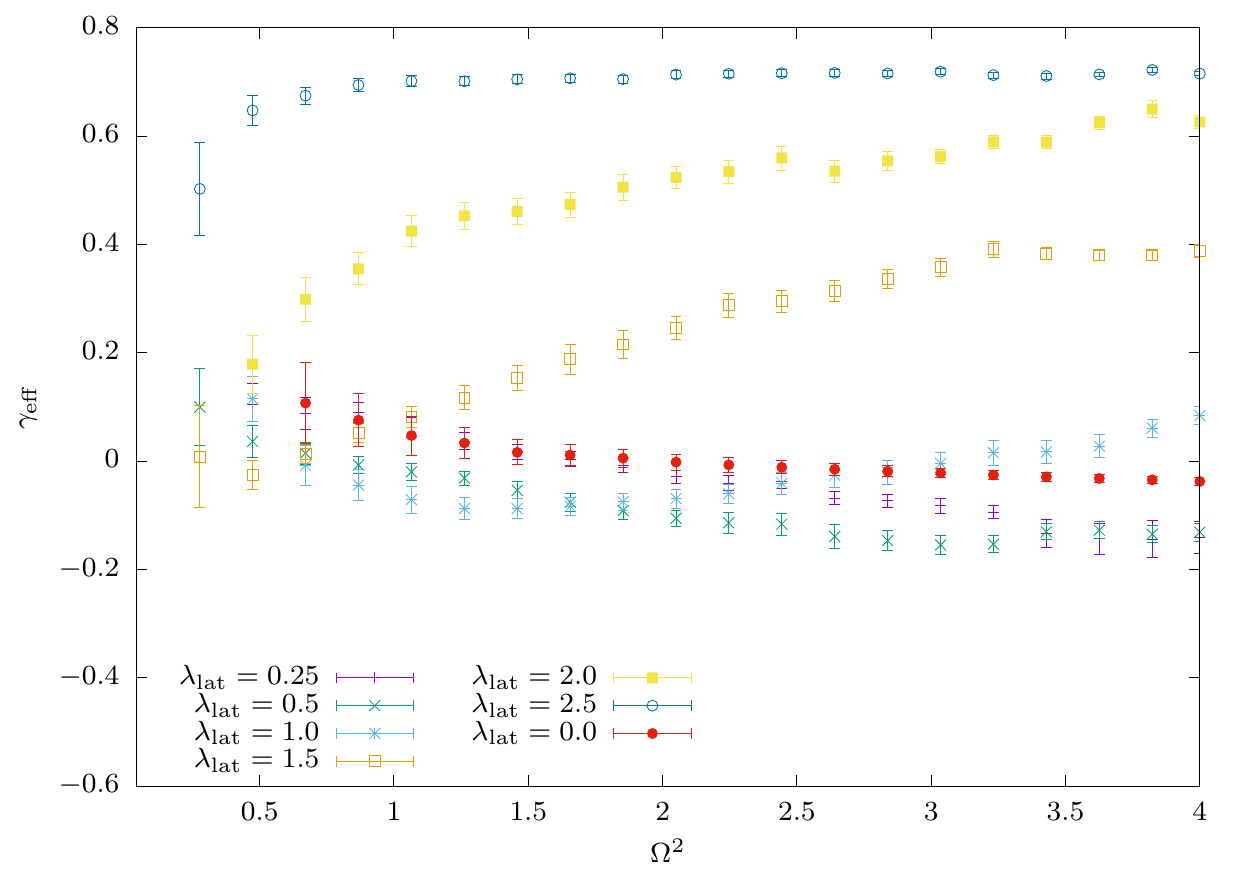}
  \caption{Effective anomalous dimension \gaEff obtained from using \eqref{eq:modenumber} to fit the stochastic Chebyshev mode number across the window $[0, \Om^2]$, for $16^4$ lattices with gauge group U(2).  Results from fitting the analytic free-theory mode number are included for comparison.  The errorbars are dominated by the standard fit error, and we omit results from fits that produce errors larger than $0.1$.}
  \label{fig:all16plainfit}
\end{figure}

We now confirm these main features through more quantitative analyses enabled by our precise data for mode number, focusing on the $L = 16$ ensembles with gauge group U(2).
As a first investigation we use \eqref{eq:modenumber} to fit the stochastic Chebyshev mode number data across the complete window $[0, \Om^2]$, obtaining the results shown in \fig{fig:all16plainfit}.
We perform correlated fits and omit from the figure results from fits with standard fit errors larger than $0.1$.
Within uncertainties the results for $\lalat \leq 2$ clearly converge to the expected anomalous dimension of zero in the IR.
The stepwise behavior of the mode number in the IR makes this the most difficult region to resolve, as we illustrate by including in \fig{fig:all16plainfit} results from fitting the analytic free-theory mode number.
While the strongest coupling $\lalat = 2.5$ still shows a trend of \gaEff approaching zero in the IR, the current data do not suffice to establish this concretely.
Either (or both) larger lattice volumes or more refined analyses are needed for this coupling.

\begin{figure}
  \centering
  \includegraphics[width=0.8\textwidth]{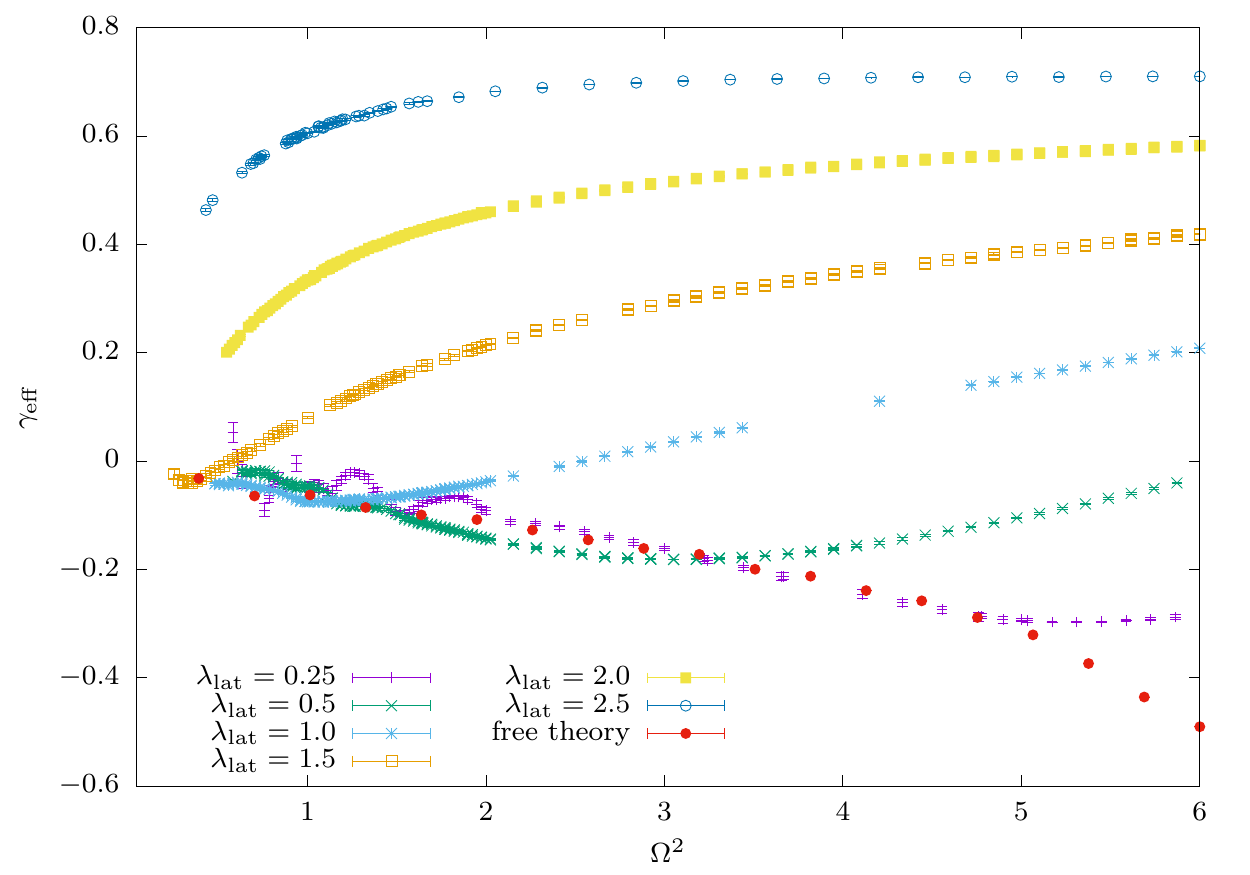}
  \caption{Effective anomalous dimension \gaEff obtained from using \eqref{eq:modenumber} to fit the stochastic Chebyshev mode number across windows $[\Om^2, \Om^2 + \ell]$ with $\Om^2 \geq 0.1$ and a fixed length $0.03 \leq \ell \leq 1$ for each $16^4$ ensemble with gauge group U(2).  We again include free-theory results (for $\ell = 1$, cf.\ \fig{fig:fitgamma}), and now omit results from any fits that produce a correlated $\chidof > 10$.}
  \label{fig:all16fit}
\end{figure}

To begin exploring alternative analyses, in \fig{fig:all16fit} we present results obtained by fitting the stochastic Chebyshev mode number data across windows $[\Om^2, \Om^2 + \ell]$ of fixed size $\ell$.
This is the procedure we presented in \secref{sec:free}, and as explained there even the free theory shows significant deviations from zero.
We have tested several possible window lengths $\ell$, excluding those that are so small they produce large oscillations in $\gaEff$, as well as those that produce large correlated $\chidof > 10$.
In \fig{fig:all16fit} we again include results from the corresponding fits of the analytic free-theory mode number with $\ell = 1$, which were previously shown in \fig{fig:fitgamma}.
Compared to \fig{fig:all16plainfit} this analysis produces much smaller uncertainties, with $\gaEff \approx 0$ in the IR for $\lalat \leq 1.5$.
There are also clear trends towards zero for $\lalat = 2$ and $2.5$, but those results remain non-vanishing down to the smallest $\Om^2$ we can access with this approach on $16^4$ lattices.

\begin{figure}
  \centering
  \includegraphics[width=0.8\textwidth]{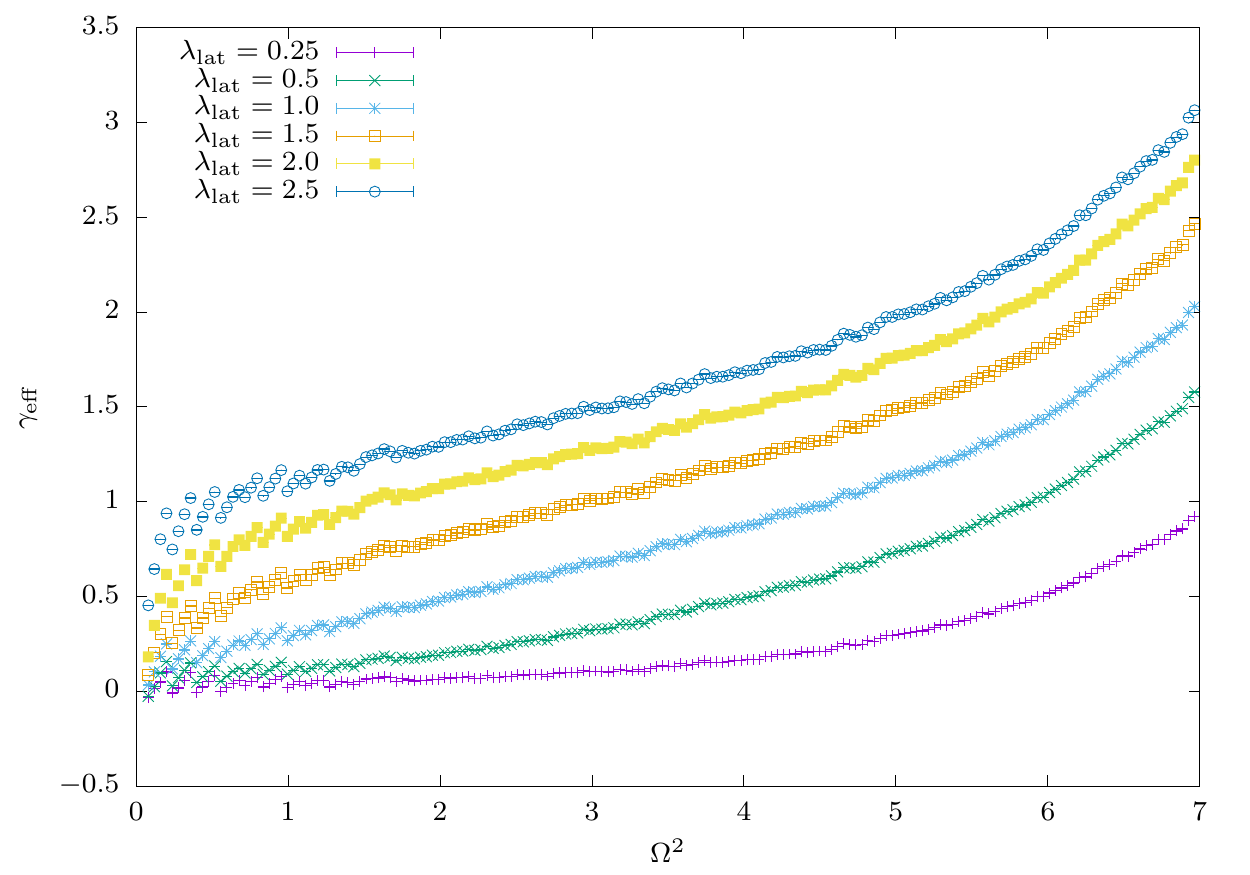}
  \caption{Effective anomalous dimension \gaEff obtained from \eqref{eq:modneff2} with reference scale $\Om_1^2 = 8$, for all six $16^4$ ensembles with gauge group U(2).}
  \label{fig:all16ratios}
\end{figure}

Finally, in \fig{fig:all16ratios} we show results from a third method, which uses \eqref{eq:modneff2} to improve stability by normalizing the stochastic Chebyshev mode number data with respect to the free theory.
We choose the reference scale $\Om_1^2 = 8$ to be beyond of the range considered in the figure without becoming too large.
In this approach oscillations from the stepwise behavior of the mode number are clearly visible for small $\Om^2$, but the main features discussed above remain the same: \gaEff increases for stronger couplings $\lalat$, while approaching zero in the IR.
Again, the $16^4$ lattice volume doesn't suffice to completely resolve the convergence to zero for the strongest couplings we consider.

\section{\label{sec:NcL}Gauge group and volume dependence}
\begin{figure}
  \centering
  \includegraphics[width=0.8\textwidth]{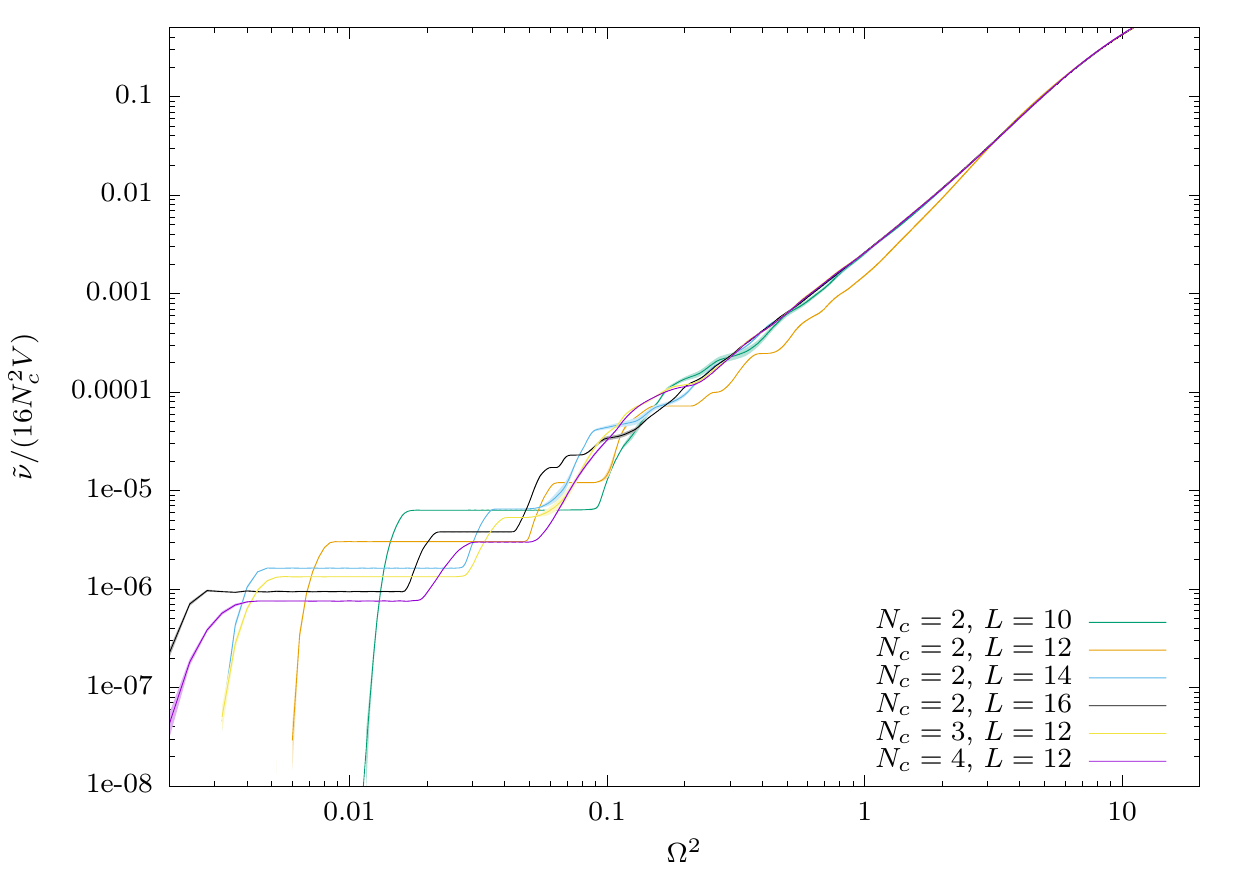}
  \caption{Normalized mode number for different volumes $L^4$ and gauge groups U($N_c$) at a fixed lattice 't~Hooft coupling $\lalat = 0.5$.}
  \label{fig:lambda050allncvol}
\end{figure}
\begin{figure}
  \subfigure[$\lalat=1$]{\includegraphics[width=0.45\textwidth]{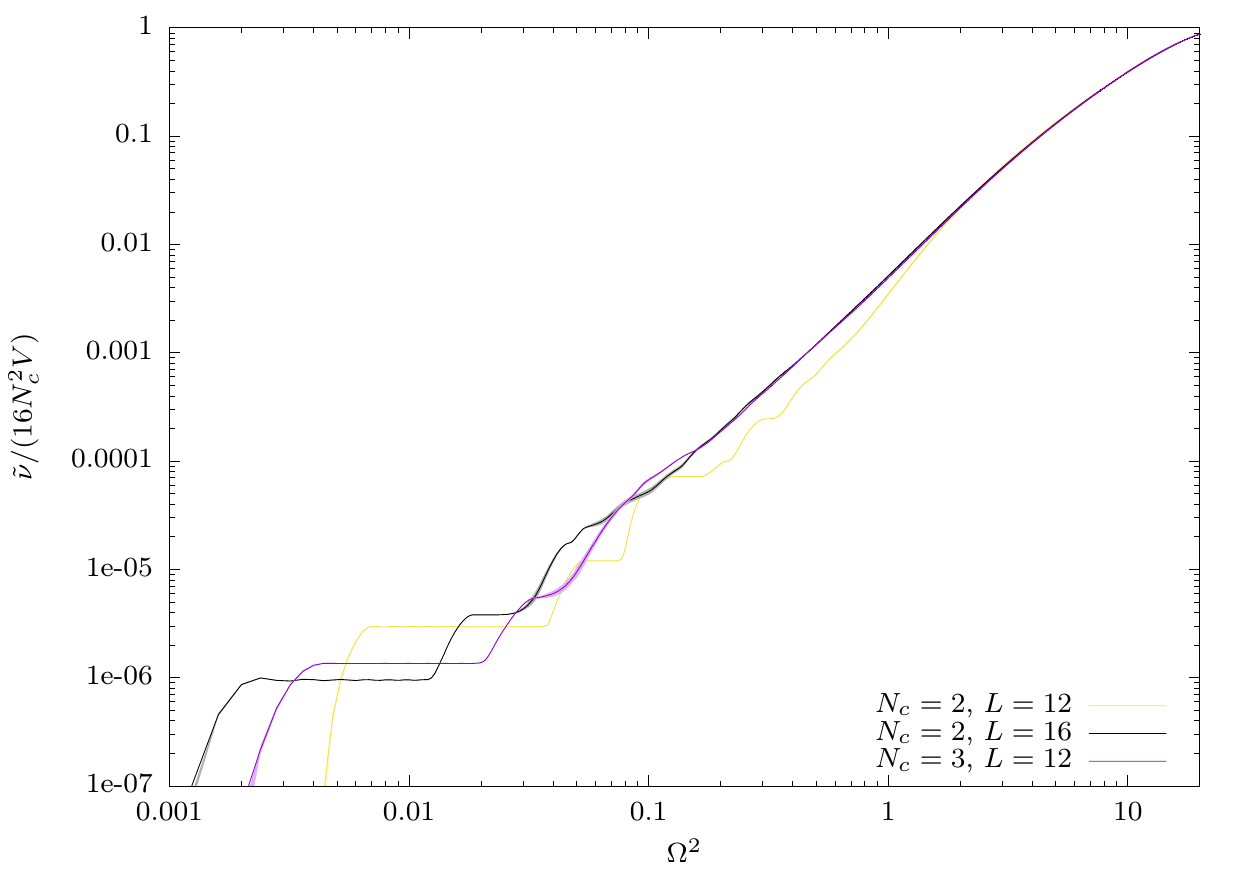}}\hfill
  \subfigure[$\lalat=1.5$]{\includegraphics[width=0.45\textwidth]{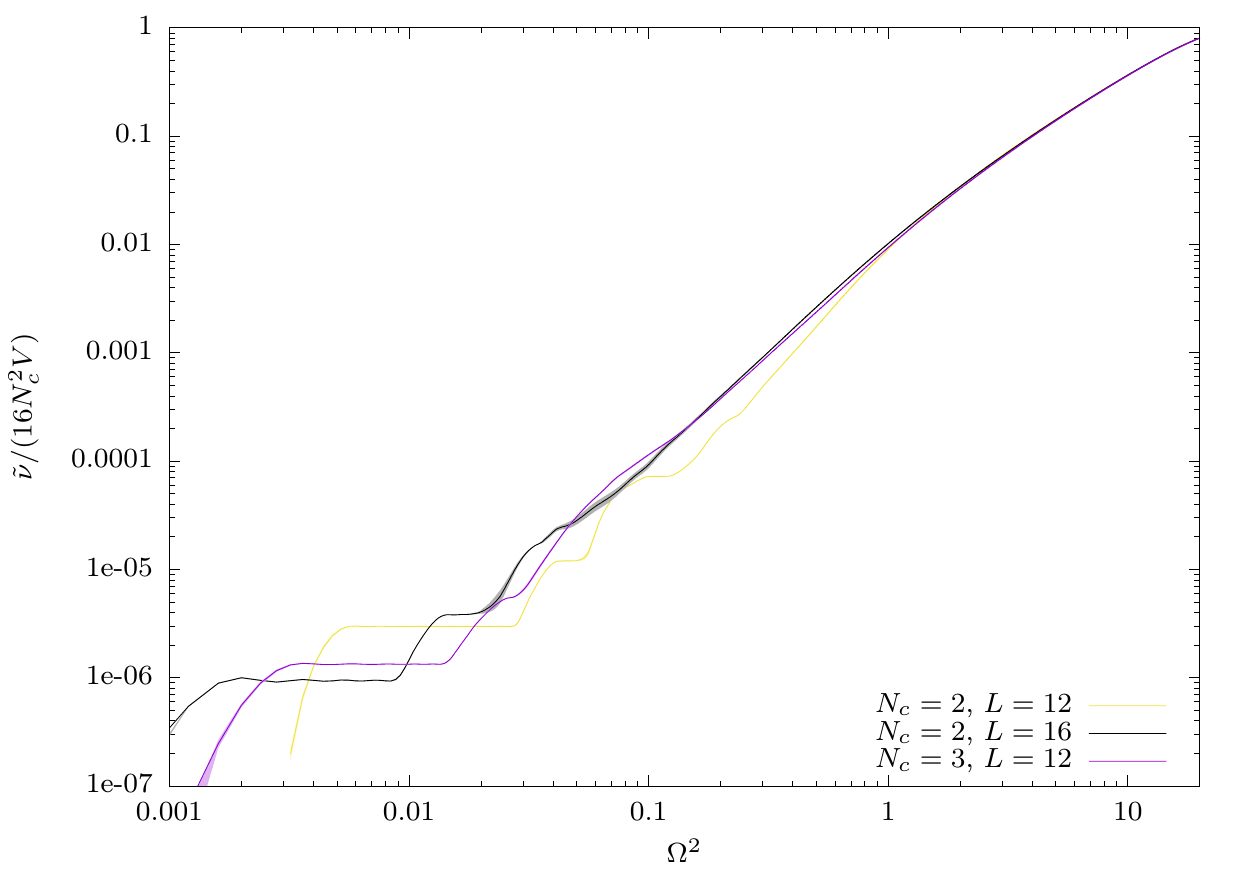}}
  \caption{Normalized mode number for different volumes $L^4$ and gauge groups U($N_c$) at couplings $\lalat = 1$ (left) and $\lalat = 1.5$ (right).}
  \label{fig:lambda100allncvol}
\end{figure}

In \secref{sec:gamma} we focused on results for the effective anomalous dimension \gaEff for the $L = 16$ ensembles with gauge group U(2) listed in \tab{tab:ensembles}.
We now investigate the dependence of our results on the lattice volume $L^4$ and the gauge group U($N_c$).
In \fig{fig:lambda050allncvol} we fix the lattice 't~Hooft coupling $\lalat = 0.5$ and observe reasonably consistent behavior in the mode number for all available $L$ and $N_c$, with similar slopes on double-logarithmic axes implying similar \gaEff for sufficiently large $\Om^2$.
The low-lying eigenvalues clearly depend on the volume and gauge group, as expected, but away from the small-$\Om^2$ region the only outlier is the U(2) ensemble with $L = 12$, which may be related to the choice of $\mu$ for this ensemble.
Figure~\ref{fig:lambda100allncvol} shows the same consistency for the stronger couplings $\lalat = 1$ and $1.5$ where we have multiple $L$ and $N_c$ to compare.

\begin{table}
  \centering
  \renewcommand\arraystretch{1.2} % Increase the height of each row
  \addtolength{\tabcolsep}{3 pt}  % Increase separation between columns
  \begin{tabular}{c|c|l|l}
    \lalat & $L$ & $y_2$    & $y_{11}$ \\
    \hline
    0.5    & 10  & 4.064(6) & 3.603(2) \\
    0.5    & 12  & 4.16(2)  & 3.725(4) \\
    0.5    & 14  & 4.28(9)  & 3.83(2)  \\
    \hline
    1.0    & 12  & 4.36(2)  & 3.76(4)  \\
    \hline
    2.0    & 12  & 5.04(2)  & 3.751(8) \\
  \end{tabular}
  \caption{Scaling dimension $y_k = 2 / (1 + \gaEff)$ obtained from \eqref{eq:evscaling} for U(2) ensembles using the corresponding $L = 16$ results for reference.  The two estimates consider the average second and eleventh eigenvalues, $\vev{\la_2^2}$ and $\vev{\la_{11}^2}$.}
  \label{tab:evscaling}
\end{table}

As shown by \eqref{eq:evscaling}, we can use the volume scaling of individual eigenvalues to obtain alternative estimates of the effective anomalous dimension.
In part because we only directly compute the $2\times 6$ lowest $\Ddag D$ eigenvalue pairs, we expect this approach to provide only very rough estimates.
Indeed, the results $y_k = 2 / (1 + \gaEff) \approx 4$ for two representative $k = 2$ and $11$ in \tab{tab:evscaling} are significantly different from the expected $y_k = 2$.
Similar results can be seen from the data in \refcite{Weir:2013zua} for the lowest eigenmodes, while the higher modes are more consistent with $y_k = 2$.
While we expect the lowest eigenmodes to be those most affected by finite-volume effects, analyzing the free-theory eigenvalues in this way produces reasonable agreement with the expected scaling, implying that the results in \tab{tab:evscaling} cannot be directly attributed to lattice artifacts.

\begin{figure}
  \subfigure[$L^2$ scaling]{\includegraphics[width=0.45\textwidth]{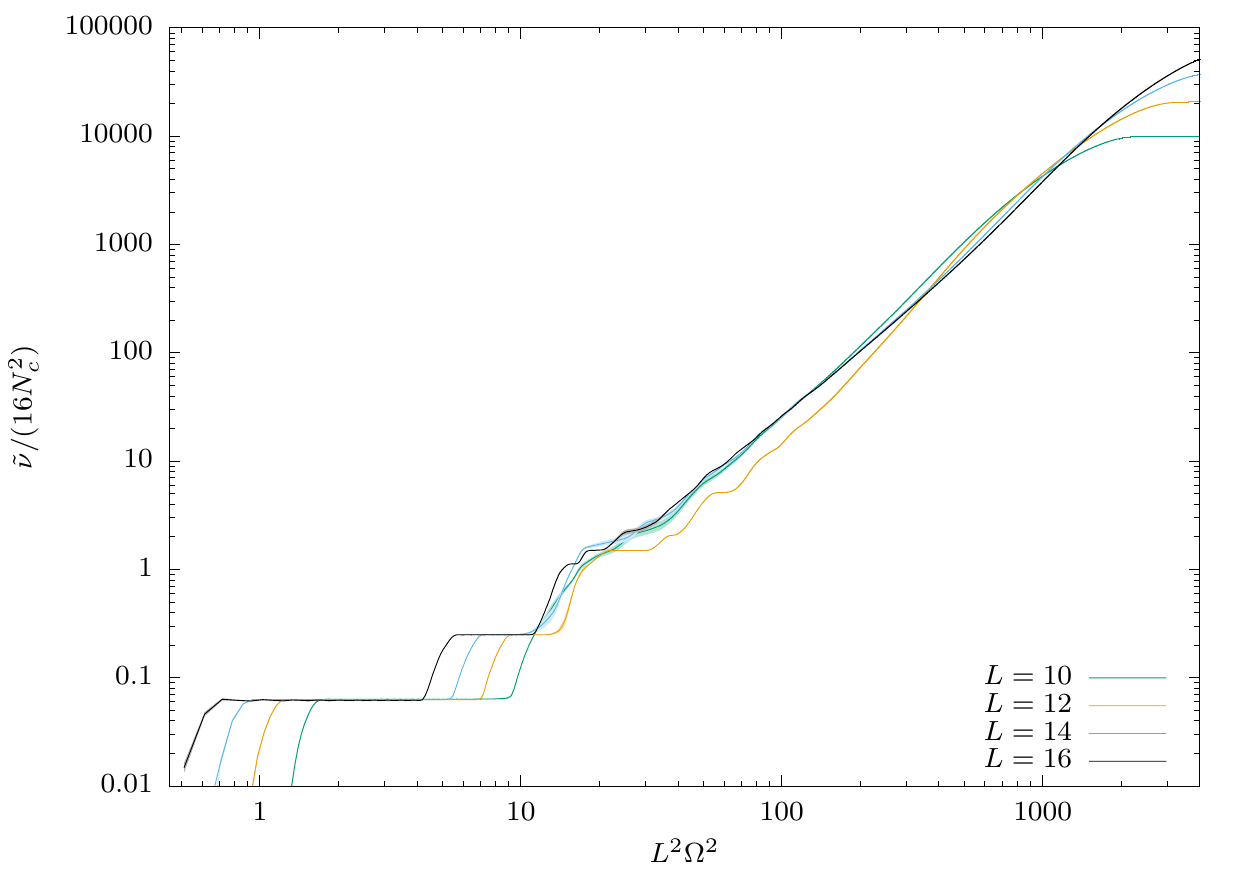}}\hfill
  \subfigure[$L^4$ scaling]{\includegraphics[width=0.45\textwidth]{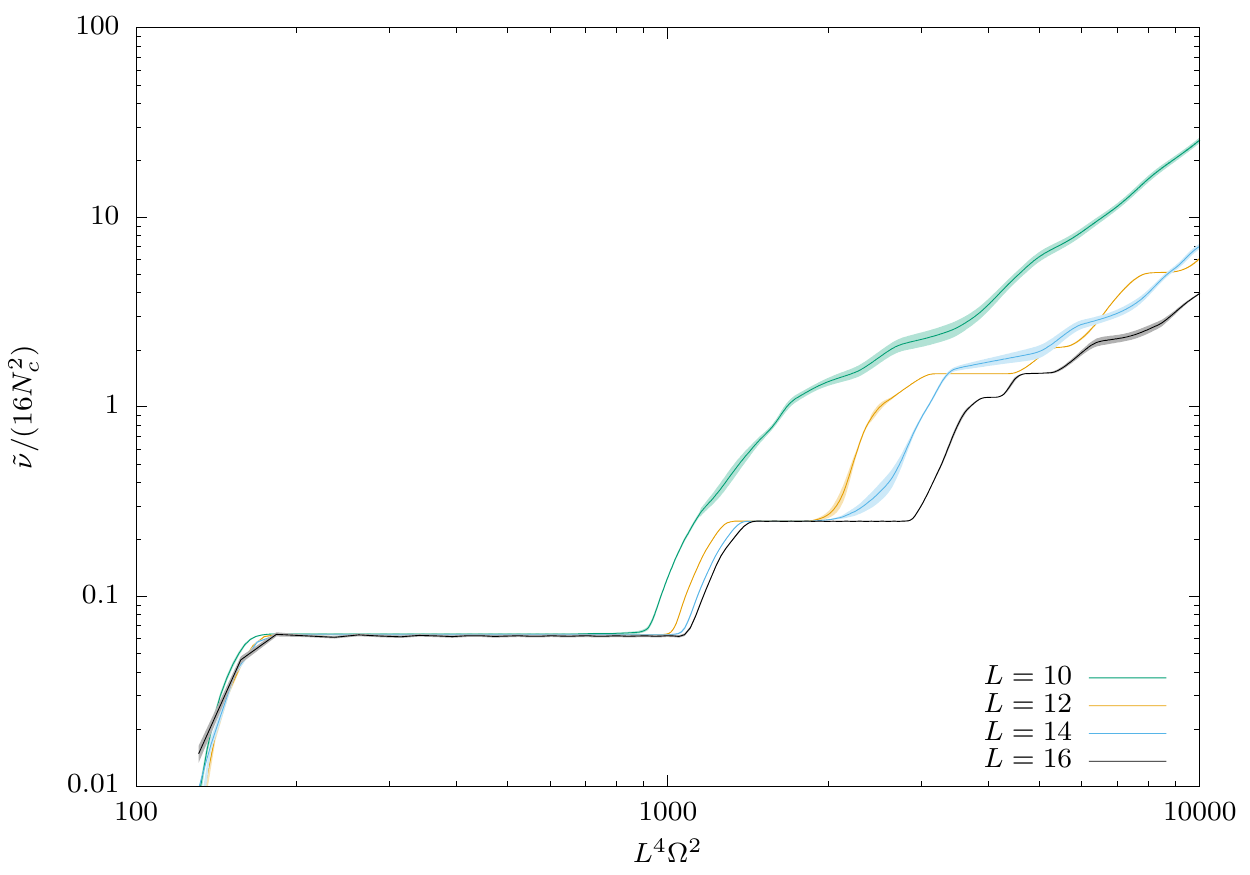}}
  \caption{Scaling of the $N_c = 2$ normalized mode number with the lattice volume $L$ at $\lalat=0.5$.  For an intermediate range of $\Om^2$ the results are consistent with $L^2$ scaling, whereas the lowest modes scale like $L^4$.}
  \label{fig:evscaling}
\end{figure}

We can obtain a more complete picture of the volume scaling by considering the stochastic Chebyshev mode number data, which we show in \fig{fig:evscaling} for fixed $\lalat = 0.5$ and $N_c = 2$.
In the left panel of this figure, we plot the normalized mode number against $L^2 \Om^2$ so that the approximate volume independence across intermediate scales corresponds to the expected $\gaEff \approx 0$.
However, the lowest modes clearly depart from this scaling, and by plotting these same data vs.\ $L^4 \Om^2$ in the right panel we can confirm that they prefer the $y \approx 4$ shown in \tab{tab:evscaling}.
It is possible that this behavior could be caused by the interactions in the theory inducing fermion (near-)zero modes despite the deformation in \eqref{eq:det} and the antiperiodic BCs we use.
The sampling of near-zero modes by the RHMC algorithm would be suppressed due to the large forces that would arise in the molecular dynamics evolution used to update the field configurations.
It may be that the interplay between the interactions in the theory vs.\ this algorithmic effect could be responsible for both the \lalat dependence of the minimum eigenvalues in \tab{tab:ensembles} as well as their unexpected volume scaling in \tab{tab:evscaling} and \fig{fig:evscaling}. % Glossing over original vs. rescaled operators

\begin{figure}
  \centering
  \includegraphics[width=0.8\textwidth]{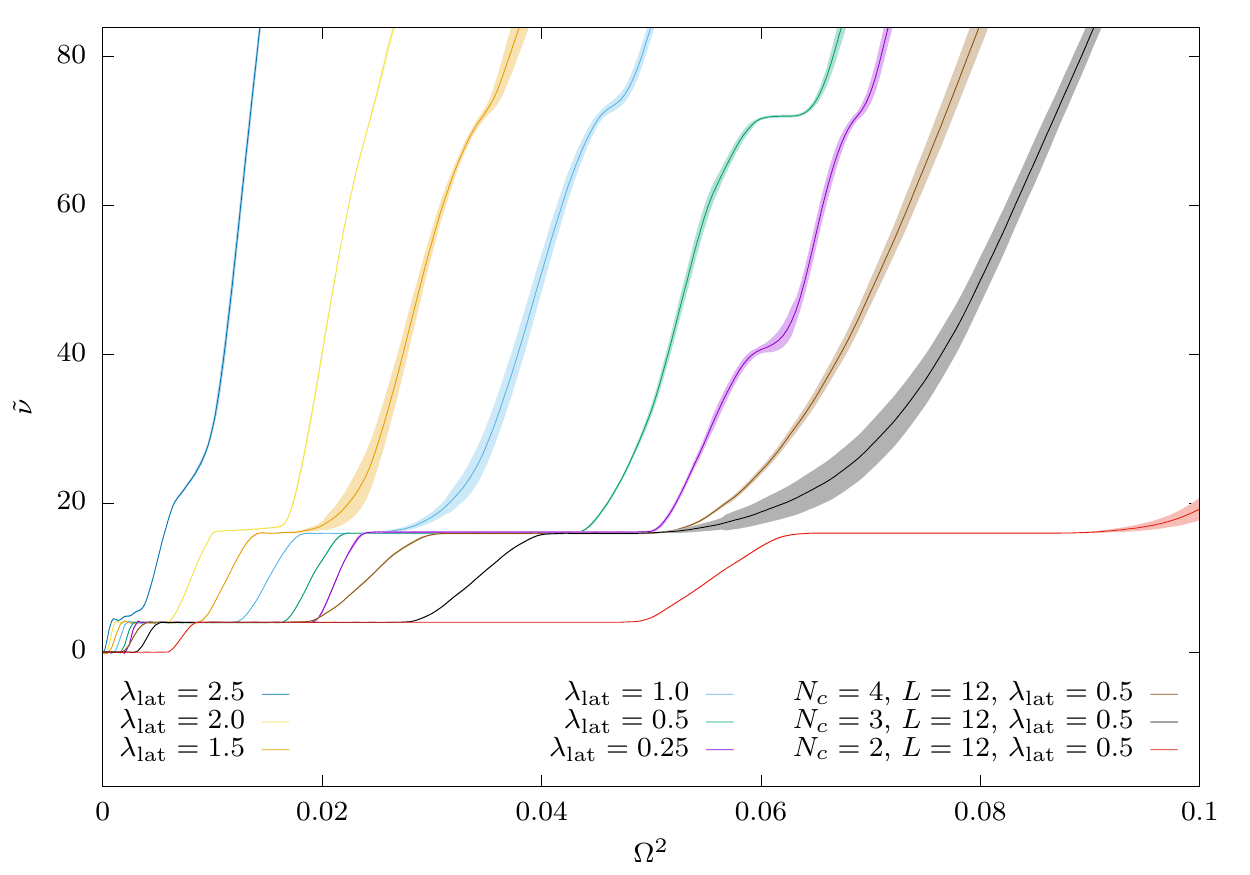}
  \caption{Total mode number in the small-eigenvalue regime ($\Om^2 \leq 0.1$) for different volumes $L^4$, gauge groups U($N_c$) and couplings $\lalat$.  If not otherwise indicated, the results are for $N_c = 2$ and $L = 16$.  The lowest modes form clusters of $4+12$ eigenvalues independent of $L$, $N_c$ and $\lalat \leq 2$.}
  \label{fig:ncscalingsmallev}
\end{figure}

We complete our discussion with a short comment on the $N_c$ dependence of our results.
In the free theory the degeneracy of the lowest eigenvalues scales with $N_c^2$, so we might expect the stepwise patterns in the mode number at small \Om to increase $\sim N_c^2$ in the interacting theory as well.
As shown in \fig{fig:ncscalingsmallev}, we do not see such behavior in our stochastic Chebyshev mode number data.
For $\lalat \leq 2$, all the lattice volumes and gauge groups we analyze exhibit a cluster of $4$ lowest modes followed by a second cluster of $12$ additional modes, well separated from the rest of the spectrum.
While the size of the lowest eigenmodes $\la_{\text{min}}^2$ scales approximately proportional to $1 / N_c$, this empirical observation might be affected by the interplay between interactions vs.\ algorithmic details discussed above.

\section{\label{sec:conc}Conclusion}
We have presented initial non-perturbative investigations of the RG properties of $\cN = 4$ SYM regularized on a space-time lattice, as part of a broader program of numerical investigations of this theory~\cite{Schaich:2018mmv}.
Employing ensembles of lattice field configurations generated using the RHMC algorithm with an improved lattice action, we stochastically estimated the Chebyshev expansion of the mode number of the fermion operator $\Ddag D$, and analyzed these data to obtain an effective anomalous dimension \gaEff that is expected to vanish in the conformal continuum theory, $\ga^*(\la) = 0$.
These RG properties are quite challenging to study in discrete lattice space-time, due to the necessary breaking of conformal invariance and 15 of the 16 supersymmetries despite the preservation of a closed supersymmetry subalgebra by our formulation of lattice $\cN = 4$ SYM.
Our work reported here provides new information about the resulting lattice artifacts and the recovery of $\cN = 4$ SYM in the continuum limit.

In addition to our main non-perturbative numerical analyses, we also considered the free theory on the $A_4^*$ lattice.
This allowed us to check our stochastic estimation of the mode number, to test our extraction of the effective anomalous dimension, and to explore the lattice artifacts that lead to non-zero \gaEff even in this case.
We carried out further validation of our main Chebyshev results by checking them against direct measurements of the low-lying eigenvalues as well as a more computationally expensive stochastic projection method. % Even though we don't show any projection method results
All three of these analyses are provided in our public parallel software.
We compared three strategies for extracting the effective anomalous dimension from the Chebyshev mode number, observing the same general features in each of the corresponding Figs.~\ref{fig:all16plainfit}--\ref{fig:all16ratios}.
These show \gaEff increasing for stronger lattice 't~Hooft couplings $\lalat$, while still approaching zero in the IR, with the convergence to zero not completely resolved for the strongest $\lalat = 2.5$ we consider, which may need to be analyzed on larger lattice volumes.

Finally, we considered the dependence of our results on the lattice volume $L^4$ and the gauge group U($N_c$).
While we observed the expected $L^2$ scaling of our stochastic mode number results in an intermediate range of scales, the lowest eigenvalues instead scaled like $L^4$, which we speculated could be connected to the RHMC algorithm used to generate lattice field configurations.
The multiplicities of those low-lying eigenvalues also don't display the expected dependence on $N_c^2$, while their size scales approximately proportional to $1 / N_c$, which may also be affected by algorithmic details.

Overall, while our numerical results indicate that lattice artifacts are increasingly significant at larger couplings $\lalat$, the convergence towards the expected $\ga^*(\la) = 0$ in the IR provides reassurance that the correct superconformal continuum limit remains accessible from $16^4$ lattice volumes for $\lalat \leq 2$.
For larger $\lalat \geq 2.5$ it seems larger lattices will be needed in order to be confident that the continuum limit is under control.
This is useful input for other ongoing studies of lattice $\cN = 4$ SYM, investigating inter alia the static potential and the Konishi operator scaling dimension~\cite{Schaich:2018mmv}.
Similar work can also be considered for alternative $\cN = 4$ SYM lattice actions currently being explored~\cite{Catterall:2020lsi}.
Our results also highlight features of the lowest-lying eigenmodes that are not yet clearly understood, and merit further consideration.

\section*{Acknowledgements}
We thank Simon Catterall and Joel Giedt for ongoing collaboration on lattice $\cN = 4$ SYM, and particularly appreciate helpful conversations with Joel concerning the mode number and mass anomalous dimension.
GB acknowledges support from the Heisenberg Programme of the Deutsche Forschungsgemeinschaft (DFG) Grant No.~BE 5942/3-1.
DS was supported by UK Research and Innovation Future Leader Fellowship {MR/S015418/1} and STFC grant {ST/T000988/1}.
Numerical calculations were carried out at the University of Liverpool, the University of Bern, and on USQCD facilities at Fermilab funded by the US Department of Energy.

\section*{Appendix: Additional information on ensembles}
In prior work~\cite{Catterall:2015ira}, the approach to the continuum limit of lattice $\cN = 4$ SYM was mainly monitored by measuring violations of Ward identities for the twisted-scalar supersymmetry $\cQ$.
These violations are introduced by the soft breaking of \cQ due to \eqref{eq:pot}, and must vanish in the continuum limit in order to recover the full symmetries of $\cN = 4$ SYM.
In particular, \refcite{Catterall:2013roa} shows how the recovery of all $16$ supersymmetries of the theory results from the restoration of \cQ combined with a set of discrete R-symmetries, subgroups of the full SU(4)$_R$.

\begin{table}
  \centering
  \renewcommand\arraystretch{1.2} % Increase the height of each row
  \addtolength{\tabcolsep}{3 pt}  % Increase separation between columns
  \begin{tabular}{c|c|l||l|l|l}
    $N_c$ & $L$ & \lalat & \eqref{eq:Ward_sB} & \eqref{eq:Ward_Bilin} & \eqref{eq:Ward_det} \\
    \hline
    2     & 10  & 0.5    & 0.00051(5)         & 0.014(1)              & 0.00323(9)          \\
    \hline
    2     & 12  & 0.25   & 0.0003(1)          & 0.005(2)              & 0.0005(2)           \\
    &     & 0.5    & 0.00039(3)         & 0.0085(7)             & 0.0020(2)           \\
    &     & 1.0    & 0.00098(7)         & 0.0197(9)             & 0.0086(4)           \\
    &     & 1.5    & 0.00168(7)         & 0.025(1)              & 0.0148(6)           \\
    &     & 2.0    & 0.00226(9)         & 0.0284(8)             & 0.0204(6)           \\
    &     & 2.5    & 0.00336(6)         & 0.042(1)              & 0.0351(8)           \\
    \hline
    2     & 14  & 0.5    & 0.00022(3)         & 0.0067(6)             & 0.00149(6)          \\
    \hline
    2     & 16  & 0.25   & 0.00016(6)         & 0.0024(8)             & 0.0004(1)           \\
    &     & 0.5    & 0.00014(4)         & 0.0054(7)             & 0.0013(2)           \\
    &     & 1.0    & 0.00045(3)         & 0.0106(5)             & 0.0046(2)           \\
    &     & 1.5    & 0.00084(6)         & 0.0152(7)             & 0.0088(4)           \\
    &     & 2.0    & 0.00139(4)         & 0.0203(5)             & 0.0145(3)           \\
    &     & 2.5    & 0.00181(3)         & 0.0239(4)             & 0.0204(3)           \\
    \hline
    3     & 12  & 0.5    & 0.00023(6)         & 0.007(1)              & 0.0015(3)           \\
    &     & 1.0    & 0.00044(5)         & 0.0089(7)             & 0.0038(3)           \\
    &     & 1.5    & 0.00100(5)         & 0.0108(7)             & 0.0063(5)           \\
    \hline
    4     & 12  & 0.5    & 0.00015(4)         & 0.0023(8)             & 0.0006(2)           \\
  \end{tabular}
  \caption{\label{tab:Ward}Additional information about the ensembles summarized in \tab{tab:ensembles}.  These violations of the three \cQ Ward identities discussed in the text contain complementary information about the approach to the continuum limit.  As expected~\cite{Catterall:2015ira}, the violations decrease as \lalat decreases, as $L$ increases, and as $N_c$ increases.}
\end{table}

In this Appendix we supplement \tab{tab:ensembles} by reporting numerical results for the violations of three \cQ Ward identities for each of the $18$ ensembles we consider.
Here we briefly describe the three Ward identities under consideration, which are discussed in detail in \refcite{Catterall:2015ira}.
Each can be expressed as the vacuum expectation value of the supersymmetry transformation of a suitable local operator, $\vev{\cQ \cO}$.
Such a local operator already appears in the $\cQ$-exact part of the lattice action shown in \eqref{eq:action}.
Because the fermion action is gaussian, this Ward identity fixes the bosonic action per lattice site to be $s_B = 9N_c^2 / 2$, and we can therefore define
\begin{equation}
  \label{eq:Ward_sB}
  W_{s_B} \equiv \frac{\left|\vev{s_B} - 4.5N_c^2\right|}{4.5N_c^2}
\end{equation}
as a normalized measure of its violation.

Another local operator was pointed out by \refcite{Catterall:2014vka}:
\begin{equation*}
  \cQ \Tr\left[\eta \sum_a \cU_a \cUbar_a\right] = \Tr{d \sum_a \cU_a \cUbar_a} - \Tr{\eta \sum_a \psi_a \cUbar_a} \equiv D - F.
\end{equation*}
Here we introduce the shorthand ``$F$'' for the second term that involves the $\eta\psi_a$ fermion bilinear and ``$D$'' for the first term that depends on the equations of motion for the bosonic auxiliary field $d$,
\begin{equation}
  \label{eq:EOM}
  d = \cDbar_a^{(-)}\cU_a + G \sum_{a \neq b} \left(\det\cP_{ab} - 1\right) \Ibb_{N_c},
\end{equation}
which are affected by the deformation in \eqref{eq:det}.
Again we define a normalized measure of the violations of this Ward identity,
\begin{equation}
  \label{eq:Ward_Bilin}
  W_{\text{Bilin}} \equiv \left|\vev{\frac{D - F}{\sqrt{D^2 + F^2}}}\right|,
\end{equation}
estimating the fermion bilinear stochastically using random gaussian noise vectors.

Finally, the presence of \eqref{eq:det} in the improved lattice action makes $\vev{\cQ \Tr\eta} = \vev{\Tr d}$ non-trivial.
(The finite-difference term in \eqref{eq:EOM} vanishes identically upon averaging over the lattice volume.)
Defining $\det\cP$ to be the average of the plaquette determinant over all orientations and lattice sites, our final Ward identity violations are simply
\begin{equation}
  \label{eq:Ward_det}
  W_{\text{det}} \equiv |\vev{\mbox{Re}\;\det\cP} - 1|,
\end{equation}
which is sensitive only to the U(1) sector of $\U{N} = \SU{N}\times \Uone$.

In \tab{tab:Ward} we collect numerical results for these three \cQ Ward identity violations from the $18$ lattice $\cN = 4$ SYM ensembles summarized in \tab{tab:ensembles}.
These results provide complementary information about the approach to the continuum limit where \cQ is restored along with all the other symmetries of the theory.
Although we include normalization factors in \eqref{eq:Ward_sB} and \eqref{eq:Ward_Bilin}, these Ward identity violations can (and clearly do) all have different overall scales.
All that matters is that they vanish in the continuum limit, and \refcite{Catterall:2015ira} found (considering $\lalat \leq 2$) that the improved action we use in this work produces effective $\cO(a)$ improvement in these continuum extrapolations.
Here we will be content to note that the Ward identity violations in \tab{tab:Ward} all systematically decrease as $L$ increases towards the continuum limit---and also as \lalat decreases or as $N_c$ increases.

\bibliographystyle{utphys}
\bibliography{paper}
\end{document}